\documentclass{emulateapj}
\usepackage{apjfonts}
\usepackage{lscape}
\usepackage{amsmath}

\newcommand{\kev}{\mbox{keV}}

\newcommand{\msun}{\mbox{$M_\odot$}}

\newcommand{\hst}{\emph{HST}}
\newcommand{\chandra}{\emph{Chandra}}

\newcommand{\rosat}{{\small ROSAT}}
\newcommand{\ergs}{\mbox{${\rm erg}~{\rm s}^{-1}$}}
\newcommand{\ergsc}{\mbox{${\rm erg}~{\rm s}^{-1}~{\rm cm}^{-2}$}}

\newcommand{\uv}{\mbox{$U\!-\!V$}}%   % U-V 
\newcommand{\vi}{\mbox{$V\!-\!I$}}%   % V-I

\newcommand{\ha}{\mbox{H$\alpha$}}

\renewcommand{\marginparwidth}{0.5in}

\shortauthors{Lugger et al.}
\shorttitle{\emph{Chandra} Sources in M30}

\begin{document}

\title{\emph{Chandra} X-ray Sources in the Collapsed-Core Globular Cluster
M30 (NGC~7099)}

\author{Phyllis M. Lugger and Haldan N. Cohn} \affil{Indiana
University, Department of Astronomy, 727 E. Third St., Bloomington, IN
47405; lugger@astro.indiana.edu}
\author{Craig O. Heinke} \affil{Northwestern University, Department of
Physics \& Astronomy, 2145 Sheridan Rd., Evanston, IL 60208}

\and

\author{Jonathan E. Grindlay and Peter D. Edmonds}
\affil{Harvard-Smithsonian Center for Astrophysics, 60 Garden St.,
Cambridge, MA 02138}

\begin{abstract}

We report the detection of six discrete, low-luminosity ($L_X <
10^{33}~\ergs$) X-ray sources, located within 12\arcsec\ of the center
of the collapsed-core globular cluster M30 (NGC~7099), and a total of
13 sources within the half-mass radius, from a 50~ksec \chandra\
ACIS-S exposure.  Three sources lie within the very small upper limit
of 1\farcs9 on the core radius.  The brightest of the three core
sources has a luminosity of $L_X\,\mbox{(0.5--6~keV)} \approx
6\times10^{32}~\ergs$ and a blackbody-like soft X-ray spectrum, which
are both consistent with it being a quiescent low-mass X-ray binary
(qLMXB)\@.  We have identified optical counterparts to four of the six
central sources and a number of the outlying sources, using deep
\emph{Hubble Space Telescope} and ground-based imaging.  While the two
proposed counterparts that lie within the core may represent chance
superpositions, the two identified central sources that lie outside of
the core have X-ray and optical properties consistent with being CVs.
Two additional sources outside of the core have possible active binary
counterparts.  We discuss the X-ray source population of M30 in light
of its collapsed-core status.

\end{abstract}  


% \\[\baselineskip]
% {\bf\today}}

\section{Introduction}

M30 (NGC~7099) is one of 21 Galactic globular clusters that show
strong evidence of having undergone core collapse
\citep{Djorgovski86,Lugger95}.  The collapsed state of the core has
been confirmed by \emph{Hubble Space Telescope} (\hst) imaging
\citep{Yanny94,Sosin97,Guhathakurta98}.  \citet{Sosin97} has placed an
upper limit of 1\farcs9 (0.08~pc) on the core radius of M30 from
high-resolution \hst\ Faint Object Camera imaging.  The
extraordinarily high central density of M30, which may exceed $\sim
10^6~\msun\,{\rm pc}^{-3}$, makes its core and the surrounding
power-law cusp one of the highest density environments in the Galaxy.
A high rate of stellar interactions is expected in the core and cusp
regions, resulting in blue straggler formation via stellar mergers,
binary formation via tidal capture and/or 3-body interactions, and
binary interactions such as exchange encounters \citep{Hut92}.  M30
shows some of the strongest evidence of stellar interactions seen in
any globular cluster.  It has a very high blue straggler frequency, a
very large bluer-inward color gradient, and a large deficit of bright
giants in the inner region \citep{Guhathakurta98,Howell00}.  Thus, the
central region of M30 is expected to be a conducive environment for
the formation of X-ray binaries.

Like the nearby collapsed collapsed-core globular cluster NGC~6397,
M30 does not contain any bright ($L_X \gtrsim 10^{35}~\ergs$) low-mass
X-ray binaries (LMXBs).  This opens the possibility of studying the
distribution of low-luminosity sources near the center of M30, which
would be hindered by the presence of bright sources.  Previous \rosat\
observations of M30 detected low-luminosity X-ray emission from the
vicinity of the core, with about 10$''$ positional accuracy
\citep{Johnston94,Verbunt01}.  Since this is comparable to the size of
the cusp region, the \rosat\ detection is consistent with either a
single source or a centrally concentrated source population.  In order
to further investigate the X-ray source population of M30, we have
obtained and analyzed a medium-depth \chandra-ACIS exposure.  As
expected, we detected a central population of X-ray sources.  We
describe the X-ray data, analysis method, and results in the following
sections.

\section{Data}

We obtained a 49.4~ksec ACIS-S\footnote{Advanced CCD Imaging
Spectrometer/S-Array} exposure of M30 on 2001 Nov 11, with the center
of the cluster placed in the S3 chip.  We chose the back-illuminated
S3 chip for its high sensitivity and spectral resolution.  The cluster
center was offset by $-0\farcm7$ in the Y-coordinate from the nominal
ACIS-S aimpoint, to ensure that the half-mass region of the cluster
was well covered by the chip.  The location of the cluster center was
approximately 2\farcm5 from the nearest chip edge.  We used the timed
exposure mode with the faint telemetry format.  The low reddening
($E(\bv)=0.03$) toward M30 results in a low neutral hydrogen column
density, $N_H \approx 1.7\times10^{20}~\mbox{cm}^{-2}$.  We have
adopted a distance $9.0\pm0.5$~kpc for M30, based on the
\emph{Hipparcos} results of \citet{Carretta00}.  Our adopted detection
threshold of 4.5 counts, for the 49.4~ksec ACIS-S exposure, translates
to a flux of $F_X \approx 4\times10^{-16}~\ergsc$ in the 0.5--6~keV
energy band, for a range of spectral models, e.g.\ a 6~keV thermal
plasma spectrum.  This limiting flux corresponds to a luminosity of
$L_X \approx 4\times10^{30}~\ergs$.

\epsscale{0.9}
\begin{figure*}[t]
\figurenum{1}
%% m30_1024_0.3-7_with_sources_rh.ps
\plotone{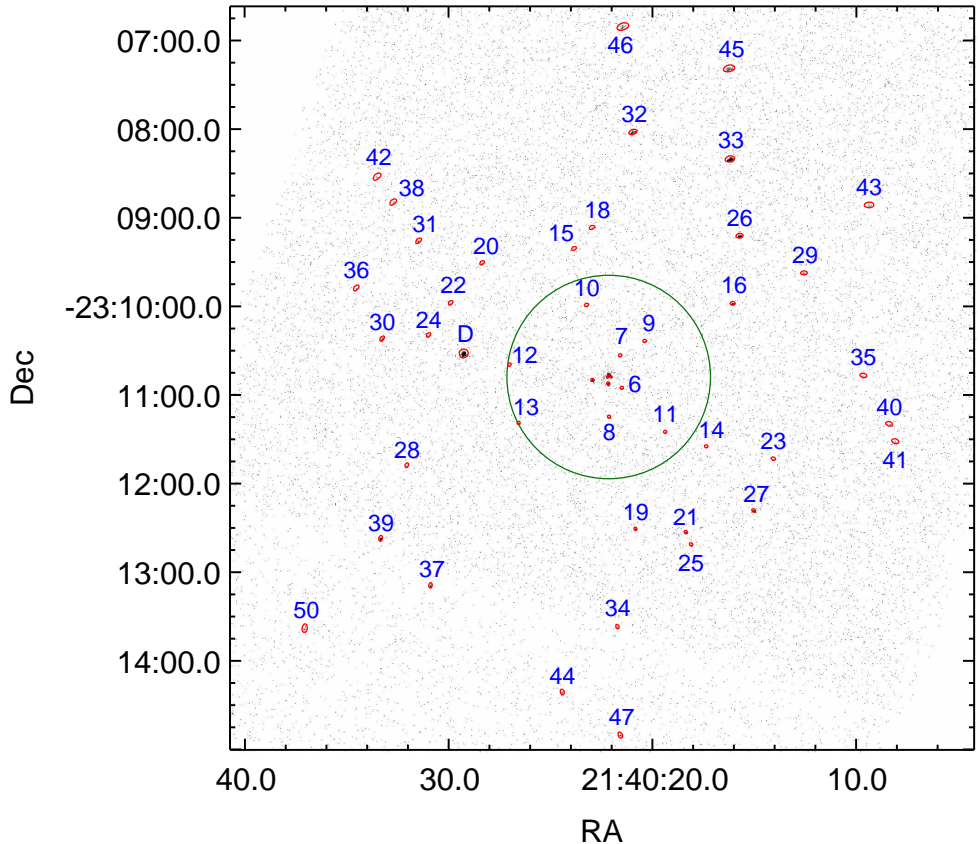}
\figcaption{\chandra\ ACIS-S image of M30 and the surrounding region
  in the 0.3--7~keV energy band.  The cluster half-mass radius of
  1\farcm15 is indicated by the circle.  The extraction regions for
  the detected sources are indicated by small polygons and are labeled
  by the source numbers from Table~\ref{tab:sources}.  The five
  central sources are labeled in Fig.~\ref{fig:acis_64_all}.  Source~D
  is the brightest object in the field.  \emph{(See the electronic
  edition of the Journal for a color version of this figure.)}
\label{fig:acis_1024_0.3-7}}
\end{figure*}

\bigskip

\section{\chandra\ Data Analysis and Results}

\subsection{Basic Processing}

We reprocessed the data from level 1 data products using the CIAO 3.2
software\footnote{http://cxc.harvard.edu/ciao/}, using current
calibration files and excluding the pixel randomization that is
performed in the standard processing to remove the artificial
substructure introduced by the precession of the spacecraft.  An
exposure of 50~ksec includes a sufficient number of dither cycles
($\sim 50$) to smooth out the substructure without additional
randomization.  The exclusion of randomization is expected to sharpen
the image slightly, which is useful in this case because of potential
source image crowding in the collapsed core region.  We used the
acis\_run\_hotpix CIAO script to identify bad columns and bad pixels,
and removed these as well as events not exhibiting one of the standard
ASCA grades.  We also applied time-dependent gain corrections and
charge transfer inefficiency corrections where appropriate.  Periods
of high background were searched for, but none was found.

\begin{figure}[t]
\figurenum{2}
%% m30_64_0.3-7_centered_with_sources_rc.ps
\plotone{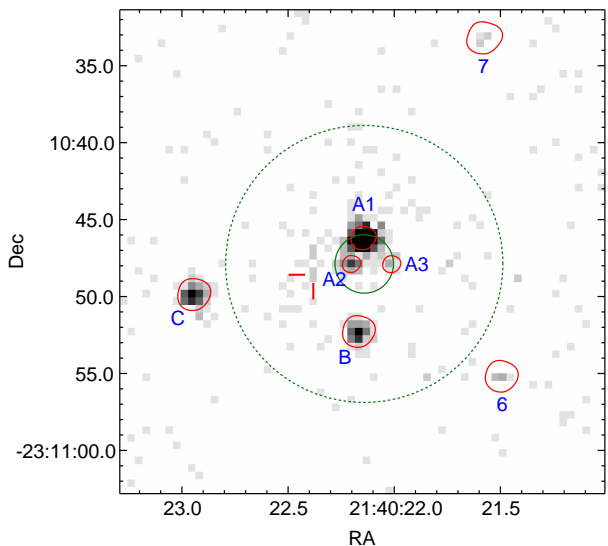}
\figcaption{ The central 32\arcsec$\times$32\arcsec\ region of the
  \chandra\ ACIS-S image of M30 in the energy band $0.3-7$~keV\@. The
  cluster core radius upper limit of 1\farcs9 is indicated by the
  central circle.  The extraction regions for the detected sources are
  indicated by polygons.  The possible millisecond pulsar X-ray
  counterpart that has been reported by \protect\citet{Ransom04} is
  indicated by the tick marks to the left of source A2.  The dashed
  circle has a radius of 9\arcsec\ = 5\,$r_c$ and is used in the
  estimation of possible unresolved flux (\S\ref{unresolved_flux}).
  \emph{(See the electronic edition of the Journal for a color version
  of this figure.)}
\label{fig:acis_64_all}}
\end{figure}

The \chandra-ACIS X-ray background is energy dependent, with a fairly
sharp rise above 7~keV\@.  Figure~\ref{fig:acis_1024_0.3-7} shows an
$8\farcm5\times8\farcm5$ region about the cluster center, in the
0.3--7~keV energy band, with the cluster half-mass radius \citep[$r_h$
= 1\farcm15;][]{Harris96} indicated and the detected sources labeled
with the numbers from Table~\ref{tab:sources}.  There is a strong
concentration of sources about the cluster center, with a decreasing
density toward the half-mass radius.  Figure~\ref{fig:acis_64_all}
shows a central 32\arcsec$\times$32\arcsec\ region of the ACIS-S image
with the standard 0\farcs5 binning.  Three bright central sources
(labeled A1, B, and C) are clearly visible and two additional fainter
sources (A2 and A3) are seen within the core.  We shall refer to the
brightest source in the field as source D\@.

\epsscale{1.2}

\begin{figure}[t]
\figurenum{3}
%% m30_64_0.3-2.5-7_centered_with_sources.ps
\plotone{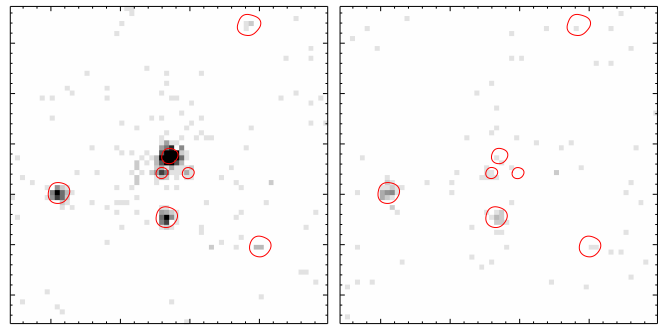}
\figcaption{A comparison of \chandra\ images of the central
  32\arcsec$\times$32\arcsec\ region of M30 in low-energy versus
  high-energy bands.  The left panel shows the $0.3-2.5$~keV band and
  the right panel shows the $2.5-7$~keV band.  The source extraction
  regions are as in Fig.~\ref{fig:acis_64_all}.  Note that the likely
  qLMXB, source A1 in Fig.~\ref{fig:acis_64_all}, nearly disappears in
  the high-energy band.  \emph{(See the electronic edition of the
  Journal for a color version of this figure.)}
\label{fig:acis_64_lo-hi}}
\end{figure}

The optical center of the cluster is located about 1\farcs8 (3.6 pix)
south of the centroid of the brightest source A1, as determined from
our boresight correction (see \S\ref{astrometry}).
Figure~\ref{fig:acis_64_lo-hi} shows a comparison of images of the
central region in low- and high-energy bands (0.3--2.5~keV versus
2.5--7~keV).  Note that source A1 nearly disappears in the high-energy
band, indicating that it has an extremely soft spectrum.  As discussed
below, this provides strong evidence that it is a quiescent low-mass
X-ray binary (qLMXB).

\subsection{Source Detection\label{source_detection}}

We used the CIAO WAVDETECT \citep{Freeman02} and PWDETECT
\citep{Damiani97} utilities for detecting sources in the 0.3--7~keV
band, after experimenting with several choices of detection bands and
source detection programs.  We used spatial scales of 1, 1.4, 2, and
2.8 pixels for both the WAVEDETECT and PWDETECT runs.  We find that
PWDETECT is very effective at identifying faint X-ray sources within a
few arcseconds of brighter sources, but appears to identify some
spurious sources and to miss real sources when applied across large
fields. WAVDETECT appears to have a more accurate algorithm for
detecting faint uncrowded X-ray sources without false detections.  The
source A2 (but not A3) is detected in the 2.5--7~keV energy band (but
not in softer or broad-band images) by WAVDETECT, while PWDETECT
detects both A2 and A3 in broad-band images.

We ran PWDETECT on the S3 chip, and WAVDETECT on a circular region
with radius 4\farcm5 centered on the cluster (i.e.\ extending out to
$3.9\,r_h$).  We also ran WAVDETECT on an image of the entire ACIS-S
detection array (6 chips), binned by a factor of two to accommodate
the increasing size of the \chandra\ point-spread function.  The
reason for extending beyond the S3 chip, in both WAVDETECT runs, was
to search for possible \chandra\ source counterparts in the cluster
halo.  For each run, we choose a significance threshold designed to
produce at most two false detections per image.

Our final source list includes the detections from the 0.3--7~keV
WAVDETECT run, removing three sources that we judge spurious (one
having less than four counts and two from spurious double detections
of off-axis sources), and adding the sources A2 and A3 identified in
the PWDETECT run.  Fifty sources were detected within 4\farcm5 of the
cluster center, and these sources are listed in
Table~\ref{tab:sources}.  An additional 46 sources were found on the
remainder of the array, after rejecting spurious sources; these
sources will be discussed elsewhere.

\subsection{Source Properties\label{source_properties}}

Table~\ref{tab:sources} (located at end of text) lists parameters for
the 50 sources within 4\farcm5 of the cluster center, including
position, distance from the cluster center (see \S\ref{astrometry}),
counts in the 0.3--7 keV band, X-ray flux, X-ray color indices, and
other identifiers.  Of these sources, 13 lie within the cluster
half-mass radius and are thus the most likely candidates to be actual
cluster members.

Source extractions were performed using the IDL script
ACIS\_EXTRACT\footnote{Version 3.65,
http://www.astro.psu.edu/xray/docs/TARA/ae\_users\_guide.html}
\citep{Broos02}, which uses CIAO and FTOOLS\footnote{Version 5.2,
http://heasarc.gsfc.nasa.gov/docs/software/ftools/ftools\_menu.html}
tools and TARA\footnote{http://www.astro.psu.edu/xray/docs/TARA/} IDL
software.  We follow the general procedures of \citet{Feigelson02}; we
briefly discuss the method here.  Point-spread functions (psfs) were
generated for each source, and polygonal extraction regions were
generated to match the 90\% encircled-energy contour (at 1.5 keV) for
most sources. For a few bright sources, we chose a larger encircled
energy; for sources A1, A2, and A3 we chose a smaller region due to
crowding.

We refined the positions of all sources by iteratively centroiding the
counts within the extraction regions.  Event lists, spectra, and light
curves were extracted for each source, and response files, background
spectra, and effective area files were constructed for each source
using current (CALDB 3.03) quantum efficiency degradation corrections
and the new MKACISRMF tool to generate responses.  The psf fraction
enclosed by the extraction regions was calculated at five energies
and interpolated for other energies, and the effective area functions
were altered to take account of these aperture corrections.

We computed background-subtracted photometry for each source in
several bands.  X-ray energy fluxes for each source were derived by
computing photon fluxes in seven relatively narrow bands.  We computed
the transformation from photon fluxes to unabsorbed energy fluxes for
each band, assuming the photoelectric absorption appropriate for the
distance to M30, for two very different spectra: a 6~keV thermal
plasma spectrum and a $10^6$~K hydrogen-atmosphere spectrum.  The
differences between the transformation coefficients were less than 4\%
for bands below 2 keV, and less than 7\% for higher energy bands.  We
adopted the 6~keV thermal plasma spectrum and summed the unabsorbed
energy fluxes (and errors) from each band to obtain the fluxes (and
errors) presented in Table~\ref{tab:sources}.

Relatively few sources in our field exhibit clear signs of
variability.  A Kolmogorov-Smirnov test (from ACIS\_EXTRACT), which was
applied to the light curve of each source, finds that only 3 of 50
sources show variability at the 99\% confidence level or higher, viz.\
sources 37, 39, and 40 (CXOU J214030.86-231308.2, 214033.30-231236.3,
and 214008.37-231118.9, respectively); all three are located beyond
3\arcmin\ from the cluster center.  Of these, source 39 has a likely
identification as a bright foreground star (see
\S\S\ref{other_sources}, \ref{other_IDs}).

\epsscale{1.0}
\begin{figure}
\figurenum{4}
%% bkgd_corr_2.eps 
\plotone{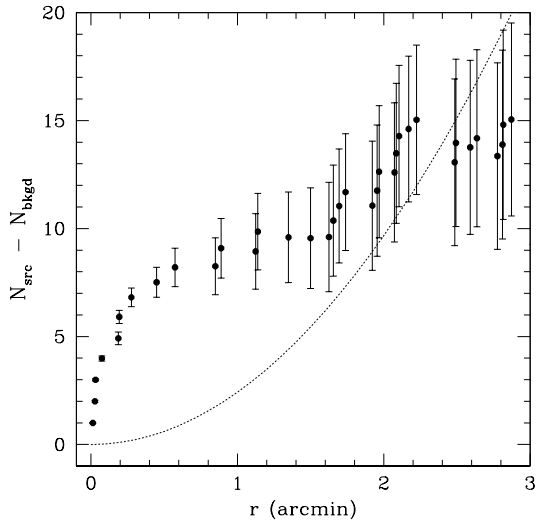}
\figcaption{Cumulative radial distribution of excess \chandra\ source
  counts over the background predicted from the \citet{Giacconi01}
  extragalactic source counts.  Error bars represent the statistical
  uncertainty of the background correction.  The dashed curve is the
  cumulative background distribution for the expected background
  density of 0.77~arcmin$^{-2}$.
\label{fig:bkgd_corr}}
\end{figure}

In order to estimate the number of sources that are likely members of
M30, we computed the expected background density of extragalactic
sources based on \citet{Giacconi01}.  Their counts predict a
background source density of $N_{\rm bkgd} = 0.77~{\rm arcmin}^{-2}$
detected above 4.5~counts in the 0.3--7~keV band, corresponding to a
flux limit $4.0\times10^{-16}~\ergsc$ in the 0.5--6~keV band for a
power-law spectrum with photon index 1.7.  Fig.~\ref{fig:bkgd_corr}
shows the cumulative radial distribution of the excess number of
sources over the background level.  The cumulative excess source
profile flattens at about 1\arcmin, in agreement with the expectation
that the source distribution should be concentrated within the
half-mass radius ($r_h = 1\farcm15$).  Fig.~\ref{fig:bkgd_corr}
indicates that the background-corrected number of cluster sources
within $r_h$ is about $10\pm2$, where the expected number of
background sources within $r_h$ is 3.2.  (We note that the error
represents the statistical uncertainty in the background correction
only.)  For comparison, \citet{Pooley03} report the detection of a
total of 7 sources within the half-mass radius of M30 to a limiting
luminosity of $4\times10^{30}~\ergs$ (0.5--6~keV; which corresponds to
a flux limit of $4\times10^{-16}~\ergsc$ for our adopted cluster
distance of 9.0~kpc), based on an analysis of the same Chandra
observations reported here.  They estimate that 1--2 of these are
background objects, using extragalactic source counts from
\citet{Giacconi01}.  We note that two of our detected sources are
located just inside of $r_h$, as seen in
Fig.~\ref{fig:acis_1024_0.3-7}, and thus may not have been counted by
\citet{Pooley03}.

It can be seen from Fig.~\ref{fig:bkgd_corr} that there is a second
rise in the excess number of sources over the expected background
between $r_h$ and $2 r_h$.  While the \citet{Giacconi01} counts
predict a total of 9.6 extragalactic sources in this region, the
actual number of detected sources is 14.  This excess persists if we
adopt a higher flux limit of $10^{-15}~\ergsc$ (0.5--6~keV), for which
4.5 sources are predicted while 9 are detected.  While this difference
does not have a high level of statistical significance, we note that
similar overdensities of sources interpreted as belonging to the
background population have been seen outside the half-mass radii of
other globular clusters \citep{Gendre03a, Heinke03c, Heinke05a},
suggesting that some of these objects may actually be X-ray sources in
the cluster halo.  Such sources might have been ejected into the halo
from the central regions or else might naturally reside in the halo
owing to their low masses.  \citet{Brandt05} have reviewed the issue
of possible cosmic variance in the extragalactic source counts,
concluding that there is some evidence for significant number
fluctuations below fluxes of about $10^{-15}~\ergsc$.  In light of
this, we also investigated by how much the background source density
would have to be increased to eliminate the apparent excess of sources
beyond $r_h$.  This would require that the density be increased by
about 30\% \citep[which is within the variance range discussed
by][]{Brandt05}, which would increase the expected background number
within $r_h$ to 4.2.  Such an upward adjustment in the background
count level would reduce the likely source population within $r_h$ by
just one source.

\bigskip

\subsection{Source Spatial Distribution}

\epsscale{1.1}
\begin{figure}
\figurenum{5}
%% m30radial_2.eps
\plotone{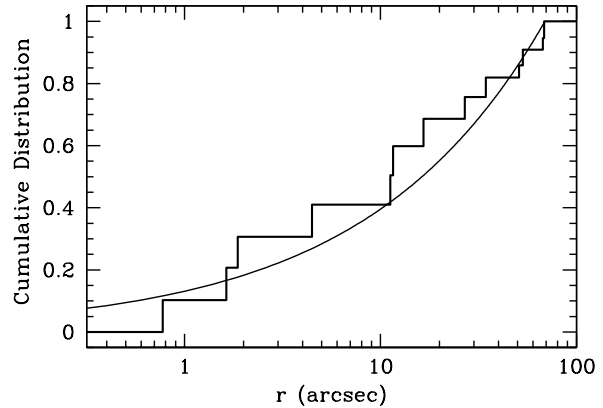}
\figcaption{Cumulative, background-corrected radial distribution of
  \chandra\ sources in M30 (stairstep) with a power-law fit (smooth
  curve).  The maximum-likelihood, power-law slope is $-1.52 \pm
  0.14$, from which we infer a mean source mass of $1.27\pm0.16~\msun$
  (see text).  \vspace*{0.2in}
\label{fig:radial}}
\end{figure}

\epsscale{1.15}
\begin{figure}
\figurenum{6}
%% CMD_M30_fx_thick.eps
\plotone{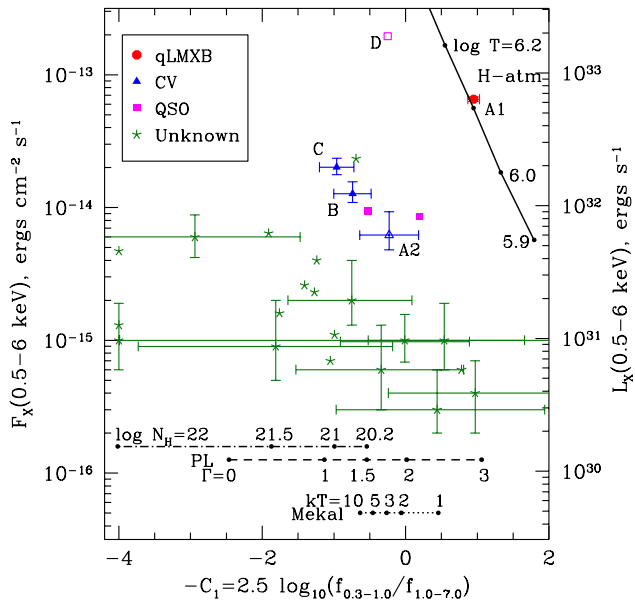}
\figcaption{X-ray color-magnitude diagram for sources within 2
  half-mass radii (2\farcm3) of the center of M30.  The abscissa is
  the negative of the index $C_1$ defined by
  Eqn.~\ref{eqn:color_index}, the left-hand ordinate is the 0.5--6~keV
  flux, and the right-hand axis gives the X-ray luminosity for an
  assumed distance of 9.0~kpc.  Sources with no soft counts are
  plotted with an X-ray color of $-C_1 = -4$.  Filled symbols are used
  for sources with a likely identification and open symbols for
  sources with a plausible identification.  Errors are plotted for
  sources lying within the 1\farcm15 half-mass radius of M30. Other
  lines indicate the colors of MEKAL thermal plasma spectra, power-law
  spectra, and intrinsically absorbed power-law spectra, or the color
  and luminosity of a 10 km NS with a hydrogen atmosphere in M30\@.
  \emph{(See the electronic edition of the Journal for a color version
  of this figure.)}
\label{fig:xcmd}}
\end{figure}

As in our previous studies of globular cluster X-ray source
distributions, we have analyzed the radial distribution of the likely
cluster sources by fitting models to the two-dimensional source
distribution using maximum-likelihood techniques
\citep[c.f.][]{Grindlay02}.  Given the small sample size ($\sim$10
sources above background within $r_h$) and the small optical core
radius of M30 ($\lesssim2\arcsec$), we have fit a pure power-law
model.  We took the background level to be that given by the
\citet{Giacconi01} extragalactic source counts, as discussed in
\S\ref{source_properties}, and use the Monte-Carlo procedure described
in \citet{Grindlay02} to correct for background.  The best-fit
power-law slope of the source surface density profile $S(r)$ is
$\alpha = d \ln S/d \ln r = -1.52 \pm 0.14$; the uncertainty
represents the 68\% range about the median value for 1000 bootstrap
resamplings of the source sample.  Figure~\ref{fig:radial} shows this
fit to the cumulative source distribution, demonstrating that it
provides a good fit to the profile.  Since M30 is a collapsed-core
cluster, the relation between cusp slope and object mass presented by
\citet{Cohn85} is appropriate; we have previously applied this
analysis to the source distribution in the collapsed-core cluster
NGC~6397 \citep{Taylor05}.  The predicted slope of a component with
mass $m$ is,
\begin{equation}
\label{eqn:cusp_slope}
\alpha = -1.89 {m \over m_d} + 0.65, 
\end{equation}
where $m_d$ is the mass of the component that dominates the cusp
potential, for which $\alpha = -1.23$.  \citet{Sosin97} has determined
a cusp slope of $-0.70\pm0.08$ for a turnoff mass group in M30.  From
this value and an assumed turnoff mass of $0.79\pm0.05\,\msun$ from
\citet[][assuming a cluster age of 14~Gyr]{Bergbusch92}, we may infer
from Eqn.~\ref{eqn:cusp_slope} that the cusp is dominated by stars
with characteristic mass $m_d=1.11\pm0.10\,\msun$; we interpret this
as representing an admixture of massive white dwarfs and neutron
stars.  Our measured slope of $-1.52\pm0.14$ for the X-ray sources then
implies a characteristic source mass of $m_X=1.27\pm0.16\,\msun$.
Given the range of uncertainty, this includes both CVs and LMXBs.

We note that the surface-brightness profile of M30 steepens
significantly from the central cusp slope beyond $\sim 0\farcm5$ from
the cluster center \citep{Lugger95}.  Thus, the measured slope for the
source population within the $1\farcm15$ half-mass radius may be best
interpreted as the mean slope for the cusp and the immediately
surrounding region.  Overall, the measured cusp slope suggests that
the characteristic X-ray source mass is either comparable to or
somewhat higher than that of the typical object that dominates the
cusp.

\subsection{X-ray Color Indices
}

\begin{figure}
\figurenum{7}
%% Color_M30_thick.eps
\plotone{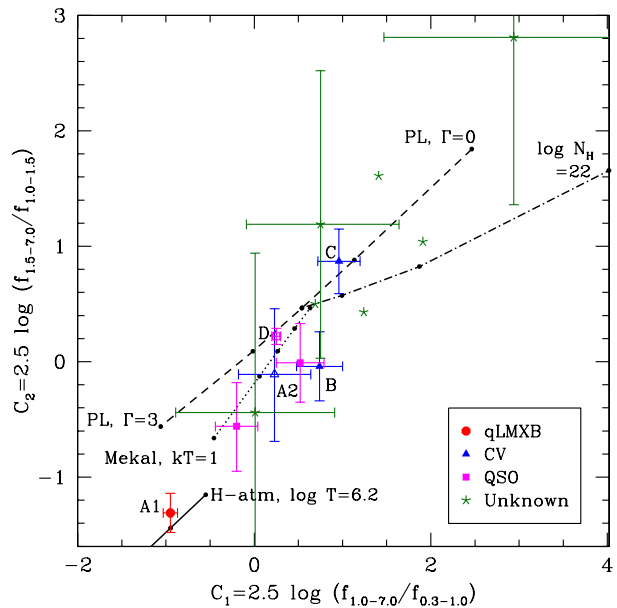}
\figcaption{X-ray color-color diagram for sources with at least 10
  total counts located within 2 half-mass radii (2\farcm3) of the
  center of M30.  The abscissa and ordinate are the indices $C_1$ and
  $C_2$ defined by Eqn.~\ref{eqn:color_index}.  Filled symbols are
  used for sources with a likely identification and open symbols for
  sources with a plausible identification.  Errors are plotted for
  sources within the half-mass radius (1\farcm15), and for the three
  objects we believe or suspect to be quasars.  The expected colors of
  various spectra are plotted, as in Fig.~\ref{fig:xcmd}.  \emph{(See
  the electronic edition of the Journal for a color version of this
  figure.)}
\label{fig:xcolor}}
\end{figure}

In order to provide some quantification of the X-ray source spectral
properties, we computed X-ray color indices following the approach of
other studies \citep[e.g.][]{Grindlay01a,Grindlay02}.  Since the
majority of detected sources are expected to be nonmembers, at a large
range of distance, a distance-independent color-color diagram provides
a useful adjunct to a color-magnitude diagram.  After some
experimentation, we defined two indices by,

\begin{subequations}\label{eqn:color_index}
\begin{eqnarray}
C_1 & = & 2.5 \log \left[ f(1.0\,\kev-7.0\,\kev) \over f(0.3\,\kev-1.0\,\kev) \right] \\[12pt]
C_2 & = & 2.5 \log \left[ f(1.5\,\kev-7.0\,\kev) \over f(1.0\,\kev-1.5\,\kev) \right].
\end{eqnarray}
\end{subequations}

\noindent The index $C_1$ is designed to be sensitive to low-energy
absorption below 1\,\kev, while $C_2$ is designed to be sensitive to
the spectral slope at higher energy.  The particular bands were chosen
to provide adequate signal-to-noise for both soft- and hard-spectrum
sources.  Figures~\ref{fig:xcmd} and \ref{fig:xcolor} show the color-flux
and color-color diagrams, respectively, for all sources within two
half-mass radii (2\farcm3) of the cluster center (only sources with
$>10$ counts are plotted in Fig.~\ref{fig:xcolor}).

We have plotted, in Figures \ref{fig:xcmd} and \ref{fig:xcolor}, the
colors where various model spectra would be expected to fall,
including a power law with photon index $\Gamma = $ 0, 1, 1.5, 2, and
3; a thermal plasma spectrum \citep[MEKAL code,][]{Phillips99}, with the low
metallicity of M30 (1.5\% solar for $\alpha$-elements) and
temperatures of 1, 2, 3, 5, and 10~keV; and a hydrogen-atmosphere
spectrum \citep[NSATMOS,][]{Rybicki05} with temperatures of $\log~T =
$ 5.9, 6.0, 6.1 and 6.2.  We have plotted both color and luminosity
for hydrogen-atmosphere neutron star spectra in Figure 6, for an
assumed 1.4~\msun, 10 km radius neutron star at a distance of 9.0~kpc.
We have also plotted the effects of increasing $N_H$ upon a power-law
model of photon index 1.5, typical of quasars or CVs that may suffer
intrinsic absorption, from the cluster standard of
$1.7\times10^{20}\,{\rm cm}^{-2}$ to $10^{21}$, $3\times10^{21}$, and
$10^{22}\,{\rm cm}^{-2}$.

As discussed in \S\ref{spectral_analysis} and \S\ref{COUNTERPARTS}, we
have identified a likely qLMXB (source A1) and two likely CVs (sources
B and C)\@. Source A1 (plotted as a filled circle) has the softest
X-ray colors of any of the objects in Figs.~\ref{fig:xcmd} and
\ref{fig:xcolor} and lies on the expected track of a 10~km neutron
star with a hydrogen atmosphere, consistent with a qLMXB
interpretation \citep{Heinke03d}.  The high flux of this object
allowed detailed spectral analysis, as described in
\S\ref{A1_spectrum}.  The two likely CVs, sources B and C, have
similar color index values, which are considerably harder than that of
source A1 and approximately consistent with a $\sim$10~keV MEKAL
spectrum or a power law of index 1.5.  The general similarity of the
colors of source A2 (plotted as an open triangle) to those of sources
B and C (filled triangles), combined with its nearly certain cluster
membership, suggests that it may also be a CV\@. We discuss its
possible positional coincidence with a moderately bright, extremely
blue star in \S\ref{source_A2} below.

The two sources that we believe are quasars, based on their optical
spectra (see \S\ref{other_IDs}), are plotted as filled squares, while
source D, which appears more likely to be a quasar than a CV based on
its optical spectrum (see \S\ref{source_D}), is plotted as an open
square.  Source D and one of the two likely quasars are
indistinguishable from CVs through their X-ray colors, while the other
likely quasar has a softer spectrum, suggestive of a photon index
between 2 and 3 or a very low-temperature MEKAL spectrum.  This
difference in X-ray color evidently indicates a difference in the
intrinsic reddening and/or the X-ray emission mechanism in these three
likely quasars.  We note that the fainter X-ray sources lying beyond
the half-mass radius (indicated by the star symbols without error bars
in Fig.~\ref{fig:xcmd}), which are probably predominantly active
galactic nuclei, show relatively harder spectra, consistent with the
results of \chandra\ extragalactic surveys \citep{Giacconi01,Kim04a}.

The similarity of the colors of likely quasars with those of likely
CVs indicates that color indices alone cannot be used to distinguish
between these source classes.  This ambiguity must be resolved by
other means, including optical identification, photometry, and
spectroscopy.

\subsection{X-ray Spectral Analysis \label{spectral_analysis}}

We fit energy spectra for the brightest sources in XSPEC, binning
the spectra to have 5, 10, 20 or more counts per bin and ignoring data
above 8~keV\@. For each fit we fix the photoelectric absorption to be
greater than or equal to the known $N_H$ column density in the
direction of M30, $1.7\times10^{20}$ cm$^{-2}$.

\subsubsection{A1, a quiescent low-mass X-ray binary \label{A1_spectrum}}

The relatively bright source A1 (CXOGLB J214022.13-231045.5) has an
X-ray color, luminosity, and spectrum similar to many quiescent
low-mass X-ray binaries identified in other globular clusters such as
NGC~6440 \citep{intzand01}, $\omega$Cen \citep{Rutledge02}, and 47~Tuc
\citep{Heinke03a} with typical $L_X = 10^{32-33}~\ergs$
\citep{Heinke03d}.  Its location within about 1\farcs8 from the center
of M30 identifies it as a certain cluster member.

The neutron stars (NSs) in qLMXBs are believed to be radiating heat
stored in their core from prior accretion episodes \citep{Brown98}, or
perhaps to be accreting at extremely low levels \citep{Campana98a}.
NSs accreting hydrogen-rich material will develop outer atmospheres of
pure hydrogen \citep[due to the rapid stratification of elements in
the atmosphere,][]{Romani87} which will determine the X-ray spectrum.
Hydrogen atmospheres shift the peak of the blackbody-like spectrum
emitted from the NS surface to higher frequencies, due to the strong
frequency dependence of free-free absorption
\citep{Rajagopal96,Zavlin96}.  \citet{Rutledge99} have shown that the
inferred radii from hydrogen-atmosphere spectral fitting of known
qLMXBs in our galaxy are consistent with the canonical 10~km NS
radius, while blackbody fits produce smaller inferred radii.

We have attempted to fit the spectrum of A1 (binned to have 20 counts
per bin) with several single-component models, each including
absorption and a (small, 2\%) pileup correction \citep[with the pileup
grade migration parameter $\alpha$ fixed to 0.5; see][]{Davis01}.
Source A1 cannot be adequately fit by an absorbed power law; the best
such fit (with a photon index of 4.9) gives $\chi^2_{\nu}=2.23$ for 30
degrees of freedom, and a null hypothesis probability of $10^{-4}$.
Bremsstrahlung or MEKAL models \citep[using an abundance of
\mbox{$[{\rm M/H}]=-1.82$} for the cluster,][]{Harris96,Carney96} give acceptable
$\chi^2_{\nu}$ values (1.15), but rather low temperatures of 0.5 keV.
Blackbody models give acceptable fits, but with rather small inferred
radii ($2.4^{+0.3}_{-0.2}$~km), and prefer a value for $N_H$ below the
cluster value.

\tabletypesize{\small}
\begin{deluxetable}{ccc}
\tablenum{2}
\label{tab:A1_spectrum}
\tablewidth{0pt}
\tablecaption{\textbf{M30 A1 Hydrogen Atmosphere Models}}
\tablehead{
\colhead{\textbf{Model Parameter}} & \colhead{\textbf{\boldmath $M$ fixed}} &
\colhead{\textbf{\boldmath $R$ fixed}} 
}
\startdata
\cutinhead{\textbf{Rybicki NSATMOS model}}
$kT$ (eV) & 94$^{+17}_{-12}$ & 124$^{+18}_{-16}$ \\[6pt]
$N_{\rm H}^{a,b}$ ($10^{20}~{\rm cm}^{-2}$) & 
2.9$^{+1.7}_{-1.2}$ & 2.9$^{+1.6}_{-1.2}$ \\[6pt]
$R_{\rm NS}$ (km) & 13.4$^{+4.3}_{-3.6}$ & (10.0) \\[6pt]
$M_{\rm NS}$ (\msun) & (1.4) &  $2.04^{+0.45}_{-0.72}$ \\[6pt]
$\chi^2_{\nu}$/dof & 1.00/30 & 1.00/30  \\[6pt]
Null hyp. prob. & 47\% & 46\%\\
\cutinhead{\textbf{Blackbody model}}
$kT$ (eV) &   191$^{+8}_{-9}$  &   \\[6pt]
$N_{\rm H}^a$ ($10^{20}~{\rm cm}^{-2}$) &    $1.7^{+0.7}_{-0}$     &    \\[6pt]
$R_{\infty}^b$ (km) & $2.4^{+0.3}_{-0.2}$ &   \\[6pt]
$\chi^2_{\nu}$/dof & 1.38/30  &  \\[6pt]
Null hyp. prob. &  7.8\%  & 
\enddata
\tablecomments{All errors are 90\% confidence limits.  Distance of
9.0 kpc is assumed.  In each column, either the mass or
the true radius of the neutron star is held fixed. \\ 
$^a$ $N_H$ restricted to be greater than the galactic column density in the
direction of M30, $1.7\times10^{20}~{\rm cm}^{-2}$. \\ 
$^b$ Reached hard limit of model.
\smallskip}
\end{deluxetable} 

\epsscale{1.1}
\begin{figure}
\figurenum{8}
%% A1_spectrum.eps
\plotone{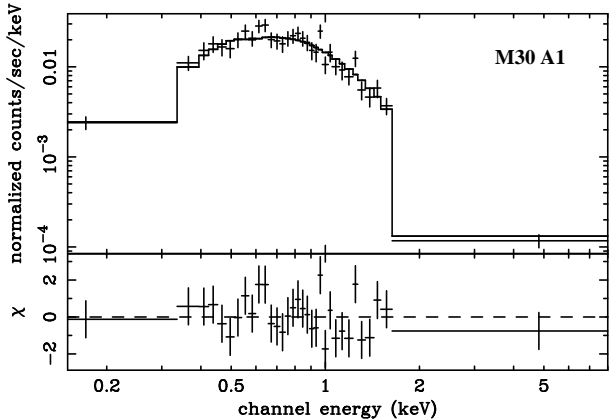}
\figcaption{Spectral fit of a neutron star, hydrogen-atmosphere model
  \protect\citep[NSATMOS,][]{Rybicki05} for source A1 (upper panel);
  deviations from the fit (lower panel).  The best-fit model
  parameters are given in Table~\ref{tab:A1_spectrum}.\bigskip
\label{fig:nsatm}}
\end{figure}

The most physically plausible model for source A1 is a neutron star
heated by accretion and radiating a blackbody-like spectrum through a
hydrogen atmosphere.  We find an excellent fit ($\chi^2_{\nu}=1.00$)
for the NSATMOS \citep{Rybicki05} model (see Fig.~\ref{fig:nsatm}),
the parameters of which are given in Table~\ref{tab:A1_spectrum}. No
additional harder component (often modeled as a power-law of photon
index 1--2 for other qLMXBs) is required; if a power-law component
with photon index 2 is included in the fit, its contribution makes up
no more than 20\% the 0.5--10~keV unabsorbed flux (90\% confidence).

The inferred unabsorbed 0.5--10~keV X-ray luminosity is
$7.1\times10^{32}~\ergs$, and the inferred bolometric luminosity is
$1.21\times10^{33}~\ergs$.  If the mass is fixed at 1.4~\msun\ and
distance is fixed at 9.0~kpc, the inferred radius is
13.4$^{+4.3}_{-3.6}$~km, consistent with a canonical 10~km NS radius
\citep[but also with the higher range found for X7 in the cluster 47
Tuc,][]{Rybicki05}.  Alternatively, with the radius constrained at
10~km, a higher mass of $2.04^{+0.45}_{-0.72}~\msun$ gives an
acceptable fit.  If neither mass nor radius is radius is fixed in the
fit, we obtain fairly broad 90\% confidence ranges of $7.0~{\rm km}
\le R \le 19.2~{\rm km}$ and $M \le 2.8~\msun$.\footnote{The NSATMOS
model considers neutron star masses above a minimum mass of
0.5~\msun.}  Fits with the NSA model \citep{Zavlin96} gave similar
results for the range of parameter space consistent with the NSA
model's single adopted surface gravity \citep[see][]{Rybicki05}.

\tabletypesize{\footnotesize}
\begin{deluxetable*}{llccccccccccc}
\tablenum{3}
\label{tab:spectral_fits}
\tablewidth{0pt}
\tablecaption{\textbf{Spectral Fits to Brighter M30 Sources} }
\tablehead{
& & \multicolumn{5}{c}{MEKAL} & & \multicolumn{5}{c}{Power-law}  \\
\cline{3-7}  \cline{9-13} \\[-5pt]
\colhead{Source} & \colhead{ID} & \colhead{$kT$} & \colhead{$N_H$} &
\colhead{$\chi^2_{\nu}$/dof} & \colhead{$F_X$} & \colhead{$L_X$} & &
\colhead{$\alpha$} & \colhead{$N_H$} & \colhead{$\chi^2_{\nu}$/dof} &
\colhead{$F_X$} & \colhead{$L_X$} \\
& & (keV) & ($10^{20}~{\rm cm}^{-2}$) & & ($10^{-15}~{\rm cgs}$) &
($10^{31}~{\rm cgs}$) & & & ($10^{20}~{\rm cm}^{-2}$) & & ($10^{-15}~{\rm cgs}$)
& ($10^{31}~{\rm cgs}$)
}
\tablecolumns{11}
\startdata  
1 & A2 &  5.0$^{+\infty}_{-1.2}$ & 1.7$^{+12.7}_{-0}$ & 1.24/5\phn & 5.8 & 5.6 &
  & 1.72$^{+1.22}_{-0.53}$ & 3.3$^{+19.6}_{-1.6}$ & 1.20/5\phn & 6.1 & 5.9 \\[6pt]
2 & A1 & 0.51$^{+0.04}_{-0.04}$ & 5.2$^{+0.9}_{-1.3}$ & 1.15/30 & 77.2 & 74.9 &
  & 4.90 & 20.5 & 2.23/30 & 169 & 163 \\[6pt]
4 & B  &  5.0$^{+12.8}_{-2.9}$ & 5.4$^{+7.8}_{-3.6}$ & 0.58/17 & 12.0 & 11.6 &
  & 1.80$^{+0.62}_{-0.40}$ & 8.7$^{+11.3}_{-5.6}$ & 0.54/17  & 12.8 & 12.3 \\[6pt]
5 & C  & 16.5$^{+63}_{-11.0}$ & 10.9$^{+9.4}_{-6.2}$ & 0.49/15 & 20.2 & 19.6 &
  & 1.46$^{+0.31}_{-0.33}$ & 13.2$^{+11.0}_{-7.9}$ & 0.49/15 & 20.7 & 20.1 \\[6pt]
16 & QSO  &  1.21$^{+0.59}_{-0.46}$ & 1.7$^{+3.4}_{-0}$ & 1.22/16 & 6.9 & 6.7 & 
  & 2.26$^{+0.52}_{-0.25}$ & 1.8$^{+5.8}_{-0.1}$ & 1.28/16 & 7.6 & 7.4 \\[6pt]
17 & D &  3.7 & 1.7 & 3.01/26 & 192 & 186 & 
  & 1.80$^{+0.06}_{-0.06}$ & 1.7$^{+0.5}_{-0}$ & 0.99/26 & 197 & 192 \\[6pt]
26 &   &  4.5$^{+2.1}_{-1.7}$ & 5.6$^{+4.8}_{-2.6}$ & 0.88/24 & 23.3 & 22.7 & 
  & 1.84$^{+0.33}_{-0.24}$ & 8.5$^{+6.2}_{-4.0}$ & 0.91/24 & 24.6 & 23.9 \\[6pt]
27 & QSO  &  5.1$^{+25.9}_{-3.3}$ & 2.1$^{+5.8}_{-0.4}$ & 0.74/14 & 9.4 & 9.1 & 
  & 1.80$^{+0.67}_{-0.45}$ & 4.8$^{+9.3}_{-3.1}$ & 0.65/14 & 9.9 & 9.6 \\[6pt]
32 &   &  3.3$^{+4.9}_{-1.7}$ & 1.7$^{+2.2}_{-0}$ & 1.00/15 & 8.8 & 8.6 &
  & 1.90$^{+0.39}_{-0.30}$ & 1.7$^{+3.8}_{-0}$ & 0.80/15 & 9.0 & 8.7 \\[6pt]
33 &   &  3.6$^{+2.0}_{-1.1}$ & 1.7$^{+0.6}_{-0}$ & 1.78/29 & 34.3 & 33.2 &
  & 1.81$^{+0.17}_{-0.14}$ & 1.7$^{+1.8}_{-0}$ & 1.51/29 & 35.4 & 34.3 \\[6pt]
39\tablenotemark{a} & star & 0.71$^{+0.13}_{-0.11}$ & 0$^{+6.7}_{-0}$ & 1.07/8\phn &
  4.4 & 4.3 &  & 2.60 & 3.3 & 2.62/8\phn & 4.2 & 4.1 \\[6pt]
45 &   &  4.0$^{+7.4}_{-2.2}$ & 2.2$^{+8.6}_{-0.5}$ & 1.15/15 & 9.2 & 9.0 & 
  & 1.85$^{+0.66}_{-0.40}$ & 4.6$^{+10.7}_{-2.9}$ & 1.16/15 & 9.8 &
  9.5 \\
\enddata
\tablecomments{Spectral fits to cluster sources, with background subtraction, in
XSPEC.  Errors are 90\% confidence for a single parameter; spectra are binned
with 20 counts/bin for $>$500 counts, 10 counts/bin for $>$200 counts, 7
counts/bin for $>$100 counts, 5 counts/bin for sources with fewer counts.  We do
not calculate errors of fit parameters for poor quality ($\chi^2_{\nu}>2.0$)
fits.  All fits include photoelectric absorption forced to be
$\geq1.7\times10^{20}~{\rm cm}^{-2}$.  X-ray fluxes in units of
$10^{-15}~\ergsc$. X-ray luminosities are in units of $10^{31}~\ergs$.}
\tablenotetext{a}{Acceptable fit requires a higher metallicity and lower $N_H$; here we use
solar metallicity.\bigskip}
\end{deluxetable*}

\subsubsection{Other cluster sources \label{other_sources}}

We fit the X-ray sources having more than 70 counts (including A1)
with MEKAL and absorbed power-law spectra; the results are summarized
in Table~\ref{tab:spectral_fits}.  Most were well-fit with MEKAL
temperatures near 5~keV or power-law indices near 1.8, with $N_H$
column densities roughly consistent with that expected toward
M30\@. Two sources are rather softer, while one is harder.

The softest spectrum (besides the qLMXB) belongs to source 39
(CXOU~J214033.30-231236.3), which we identify with a bright foreground
star below (\S\ref{other_IDs}).  This spectrum is poorly fit by a
power-law (due to significant emission lines at 0.8--0.9~keV), giving
a photon index of 2.6, and a reduced $\chi^2$ of 2.6.  Fitting with a
MEKAL spectrum of solar abundance provides acceptable fits
$\chi_{\nu}^2\sim1.1$, for a kT of $0.7_{-0.11}^{+0.13}$ keV; there is
some evidence for an additional low-temperature component as well.

The third softest spectrum belongs to one of the optically identified
quasars (see \S\ref{other_IDs}), source 16 (CXOU~J214016.05-230957.3).
This source can be fit with a power-law of photon index
$2.26_{-0.25}^{+0.52}$ or a MEKAL spectrum of kT=$1.2_{-0.46}^{+0.59}$
keV.

The hardest spectrum (with substantial counts) belongs to source C
(CXOGLB J214022.92-231049.2), which we optically identify as a CV
(\S\ref{source_C}).  This source is fit with a power-law of photon
index 1.46$_{-0.33}^{+0.31}$, or a MEKAL spectrum of
kT=$16^{+63}_{-11}$~keV.  Both require enhanced $N_H$ over the cluster
value; for the MEKAL spectrum this is $11_{-6}^{+9}\times10^{20}~{\rm
cm}^{-2}$.  This is consistent with its position in the X-ray
color-color diagram, which suggests increased $N_H$.

Finally, source D (CXOU J214029.23-231031.3), while well-fit by a
power-law of photon index $1.80_{-0.06}^{+0.06}$, is poorly fit by a
single MEKAL spectrum ($\chi_{\nu}^2=3.0$).

\subsection{Unresolved Flux \label{unresolved_flux}}

Examination of Fig.~\ref{fig:acis_64_all} suggests a possible
unresolved flux excess within the central region, e.g.\ within the
dashed circle, which has a radius of $9\arcsec = 5\,r_c$.  To
investigate this, we computed the excess counts within 9\arcsec\ of
the center of M30 by subtracting from the total counts within this
circle the combination of the counts from all detected sources within
their extraction regions, the estimated count contribution from PSF
``spill'' beyond the extraction regions, and the expected background
counts.  The resulting flux excess is $55\pm10$ counts (0.3--7\,keV).
This suggests that the presence of numerous additional faint sources
below our detection limit of 4.5~counts.  Assuming a luminosity
function in the range observed for globular clusters by
\citet{Pooley02b}, we find that the 55 excess counts within 9\arcsec\
indicate that 4--8 additional sources may be detected within this
region by improving the detection limit by a factor of 2.5.  One such
faint source is the 4-count millisecond pulsar candidate suggested by
\citet{Ransom04}, which lies just below our detection limit (see
Fig.~\ref{fig:acis_64_all}).

\section{Search for Optical Counterparts \label{COUNTERPARTS}}

\subsection{Astrometric Comparison \label{astrometry}}

The center of M30 has been extensively imaged by the \hst\ Wide-Field
Planetary Camera 2 (WFPC2).  Program GO-7379 (PI: Edmonds) has
produced particularly deep, dithered imaging in F336W ($U$), F555W
($V$), and F814W ($I$).  In order to effectively search for optical
counterparts, it is necessary to establish a common astrometric zero
point for the \chandra\ and \hst\ datasets.  While the images from
both data sets contain World Coordinate System (WCS) solutions, the
absolute accuracy of each of these is no better than about 0\farcs7.

In order to establish a common zero point, we first determined an
astrometric solution for a ground-based CTIO 1.5~m telescope $V$-band
image of M30 (kindly provided by M.~Bolte), relative to the \hst\
Guide Star Catalog 2.2 astrometric system.  We then selected
secondary astrometric standards from the ground-based image and used
these to obtain new astrometric solutions for each WFPC2 image that we
used.  This produced zero point shifts of about $0\farcs6$ relative to
the \hst\ WCS solutions given in the image headers.  Curiously, we
find a considerably larger shift for the relatively isolated
``astrometric reference star'' defined by \citet{Guhathakurta98}; our
revised position for this star is:
\begin{equation}
\label{eqn:astrometric}
\alpha =21^{\mathrm h}~40^{\mathrm m}~22\fs29,~~ 
\delta = -23\arcdeg~10\arcmin~39\farcs6 ~~\mbox{(J2000)}.
\end{equation}
This corresponds to shifts of $\Delta\alpha = -1\farcs5$ and
$\Delta\delta = +2\farcs8$ relative to the position given by
\citet{Guhathakurta98}. We note that the center position for M30 given
by \citet{Sosin97}, based on \hst\ Faint Object Camera observations,
differs from that of \citet{Guhathakurta98} by similar shifts of
$\Delta\alpha = -1\farcs1$ and $\Delta\delta = +2\farcs9$, even though
the finding charts given in these two papers refer to nearly the same
location on the sky.  In comparing WFPC2 images of M30 from different
epochs, we find that the apparent position of the astrometric star
differs by up to about 0\farcs8, which is within the expected
uncertainty of the WFPC2 image header WCS solution zero points.  Thus,
it is not clear why the shift between the present results and those of
\citet{Guhathakurta98} is nearly 3\arcsec\ in declination.

Following the completion of our astrometric analysis, we learned of an
independent analysis by \citet{Ransom04}, which is based on the
International Celestial Reference System (ICRS)\@. These authors
similarly find a substantial offset for the \citet{Guhathakurta98}
reference star.  \citet{Ransom04} report an absolute position for this
star of $\alpha =21^{\mathrm h}~40^{\mathrm m}~22\fs314,~~ \delta =
-23\arcdeg~10\arcmin~40\farcs10 ~~\mbox{(J2000)}$.  Our position is
offset from this by $\Delta\alpha = -0\farcs3$ and $\Delta\delta =
+0\farcs5$.  These offsets may largely represent the uncertainty in
our astrometric solution for the CTIO 1.5~m frame, which is based on
the \hst\ Guide Star Catalog 2.2 astrometric system.  Since we have
applied a boresight correction to put the Chandra positions on our
optical astrometric system, as described below, our astrometric
comparison of \hst\ and \chandra\ positions is not affected by any
absolute offset of our astrometry from the ICRS\@.

We next turn to the astrometric alignment of the Chandra image.
Source D, the brightest source in the \chandra\ field, is located
1\farcm6 east of the cluster center (see
Fig.~\ref{fig:acis_1024_0.3-7}) and is on the WF3 chip of the GO-7379
dataset. There is a $V\approx 20$ star with a pronounced UV excess
located within 0\farcs7 of the nominal \chandra\ position.  Adopting
this star as the likely optical counterpart to source D provides a
boresight correction to the nominal \chandra\ positions of
$\Delta\alpha = -0\farcs2, \Delta\delta = +0\farcs7$.  Based on our
subsequent identifications of counterparts to sources B and C, we
judge the accuracy of this boresight correction to be about 0\farcs1
in each coordinate.

\epsscale{1.0}
\begin{figure*}
%\rotate
\figurenum{9}
%% m30_u_pc_driz_big_with_sources_alt_rgb.ps
\plotone{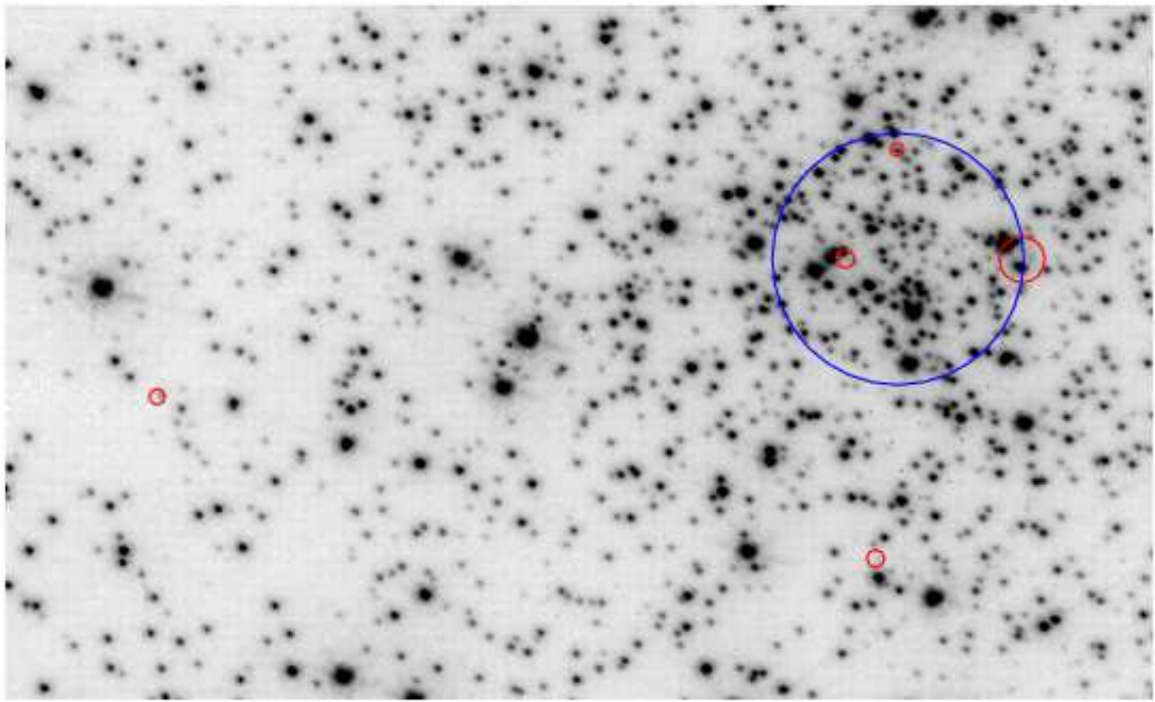}
\figcaption{A $17\arcsec\times11\arcsec$ region of the
  drizzle-reconstructed $U$-band WFPC2 Planetary Camera image of
  M30\@.  North is up and east is to the left.  The core radius is
  indicated by the large circle in the northwest quadrant.  Four-sigma
  X-ray error circles are indicated for sources A1, A2, A3, B, and C;
  see Fig.~\ref{fig:acis_64_all} for source map.  \emph{(See the
  electronic edition of the Journal for a color version of this
  figure.)\medskip}
\label{fig:u_pc_all}}
\end{figure*}

After applying this boresight correction to the \chandra\ detection
positions, to put them on our optical astrometric system, we searched
the locations of the other \chandra\ sources in the \hst\ frames.  To
facilitate this search in the crowded core region, we used the
IRAF/STSDAS\footnote{IRAF is distributed by NOAO, which is operated by
AURA, under cooperative agreement with NSF\@.  STSDAS is a product of
the STScI, which is operated by AURA for NASA\@.}  \emph{drizzle} tool
to reconstruct the Planetary Camera (PC) images from the GO-7379
dataset with $2\times$ oversampling, using the 12 separate dither
positions for the $U$, $V$, and $I$ filters.  We combined all of the
images for each filter at each position, using an exposure-weighted
average.  A portion of the resulting drizzle-reconstructed $U$-band
image is shown in Fig.~\ref{fig:u_pc_all}, with the cluster core
radius and error circles for sources A1, A2, A3, B, and C overlaid.
We also carried out a drizzle reconstruction of the entire WFPC2
fields for the GO-7379 dataset, after first using the IRAF/STSDAS
\emph{wmosaic} tool to combine the four separate WFPC2 frames into a
single mosaic for each dither position and filter.

\epsscale{1.3}
\begin{figure}
\figurenum{10}
%% m30_pc_cmd2_srcb.eps
\hspace*{-0.2in}\plotone{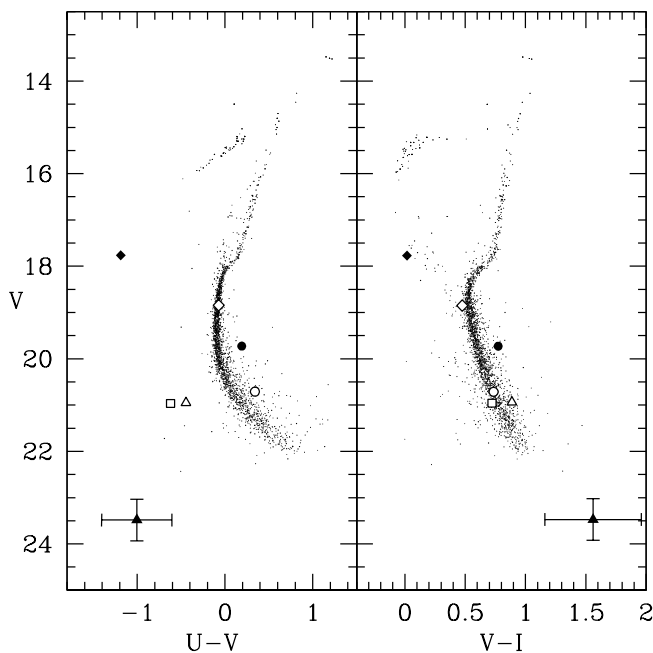}
\figcaption{CMDs for the PC frames from aperture photometry of
  drizzled images.  Points are plotted for stars detected in all three
  bands, $U$, $V$, and $I$\@.  Chandra source candidate counterparts
  are indicated as follows: source A1---open diamond; source
  A2---solid diamond; source B---solid triangle with error bars;
  source C---open triangle; source D---square; source 7---open circle;
  source 10---solid circle.  All objects lie in the PC field, except
  for source D (WF3) and source 10 (WF2).
\label{fig:pc_cmd}}
\end{figure}

High quality photometry for M30 is available from several sources,
including a \hst\ WFPC2-based $UBV$ analysis \citep{Guhathakurta98}
and a ground-based $VI$ analysis \citep{Sandquist99}.  While both
photometry sets have proved useful in the search for optical
counterparts, each has limitations.  The ground-based photometry is
most useful for the region outside of the half-mass radius, where
image crowding is much less severe than within the central region.
The \citet{Guhathakurta98} analysis is based on the undithered GO-5324
dataset and thus does not reach as deep as is possible with the
dithered GO-7379 dataset.  \citet{Pietrukowicz04} have analyzed the
GO-7379 dataset to search for variable stars in M30, but have not
published the detailed photometry.  We found it useful to perform
aperture photometry on the GO-7379 frames in order to obtain
approximate results for a greater set of candidate counterparts than
is available in the previously published photometry.
Figure~\ref{fig:pc_cmd} shows the color-magnitude diagrams (CMDs) that
result from our photometry for the PC field.

\renewcommand{\arraystretch}{1.2}
\tabletypesize{\small}
\begin{deluxetable}{lcccc}
\tablenum{4}
\label{tab:counterparts}
\tablecolumns{5}
%\tablewidth{0pt}
\tablecaption{\textbf{Candidate Optical Counterpart Photometry}}
\tablehead{
\colhead{Object} & 
\colhead{$U_{336}$} & 
\colhead{$V_{555}$} &
\colhead{$I_{814}$} 
} 
\startdata
A1 & 18.78 &     18.85     & 18.37 \\
A2 & 16.58 &     17.77     & 17.75 \\
B  & 22.5\phn &  23.5\phn  & 21.9\phn \\
C  & 20.50 &     20.94     & 20.05 \\
D  & 20.34 &     20.96     & 20.24 \\
7  & 21.05 &     20.71     & 19.98 \\
10 & 19.92 &     19.73     & 18.95 
\enddata
%
%\tablecomments{}
\end{deluxetable}
\renewcommand{\arraystretch}{1.0}

\epsscale{1.0}
\begin{figure*}
%\rotate
\figurenum{11}
%% m30_source_abc_stamps_2.eps
\plotone{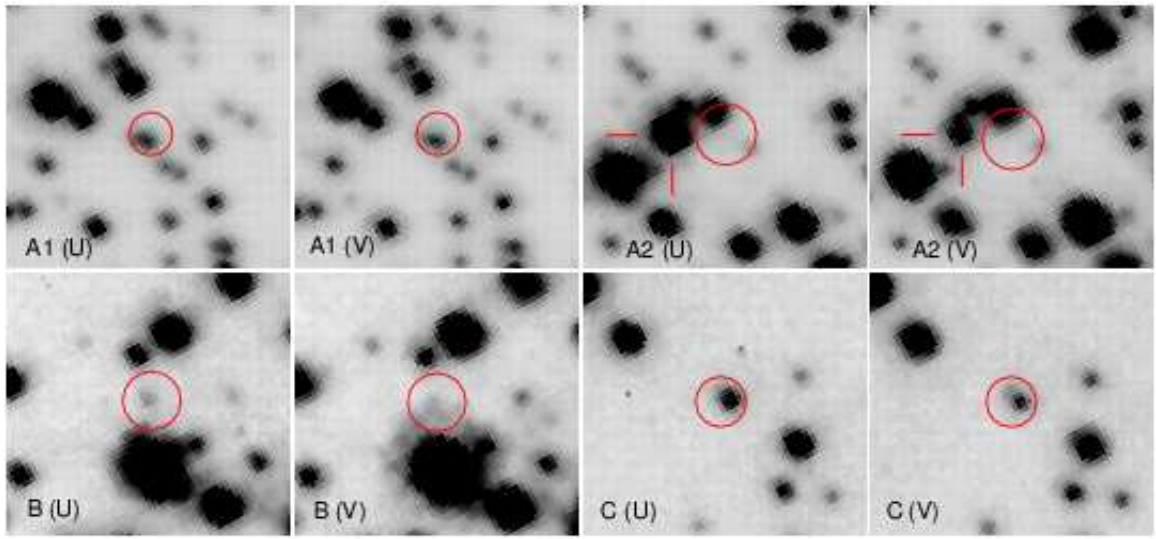}
\figcaption{Finding charts for X-ray sources A1, A2, B, and C\@. North
  is up and east is to the left.  Each chart is 1\farcs3 in east-west
  dimension.  For each source the left member of each chart pair shows
  the $U$-band image and the right member shows the $V$-band image.
  The error circles represent 4$\sigma$ uncertainties in the
  boresight-corrected source positions.  The lines in the A2 charts
  indicate the extremely blue star that lies just outside of the error
  circle.  \emph{(See the electronic edition of the Journal for a
  color version of this figure.)}
\label{fig:u_pc_stamps}}
\end{figure*}

The photometry was performed on the drizzle-reconstructed $U$, $V$,
and $I$ frames, using a 4-pixel radius aperture, which corresponds to
2 PC pixels (0\farcs09).  Stars were first detected in the $U$ frame
and then these positions were shifted to the $V$ and $I$ frames and
recentered.  Only those stars with locations within 0.5 pix of the
mean shifts are included in these CMDs, in order to filter out stars
for which blended images invalidate the aperture photometry.  A few
additional outliers were filtered out by rejecting stars that have
brighter detected neighbors within 0\farcs22.  The magnitudes were
transformed to the Johnson-Cousins system, using the calibrations
given by \citet{Holtzman95}, with aperture corrections to a fiducial
aperture of radius 0\farcs5.  Comparison of ($V$, \vi) CMD with that of
\citet{Sandquist99} indicated very good agreement of the fiducial
sequences, after additional offsets of $\Delta V = -0.1$ and
$\Delta(\vi) = +0.1$ were applied to our magnitudes.  A similar
comparison of our ($V$, \uv) CMD against that of \citet{Guhathakurta98}
similarly indicated very good agreement after applying an offset of
$\Delta(\uv) = -0.3$.  (We note that they used transformations based
on $U$, $B$, and $V$ magnitudes, while we only had $U$ and $V$
magnitudes available.)  These offsets have been adopted in the
photometric results reported below.  It can be seen, from
Fig.~\ref{fig:pc_cmd}, that the CMDs produced by this aperture
photometry process show well-defined sequences to the detection limit
(defined by the $U$-band image) of about 3 magnitudes below the
main-sequence turnoff (MSTO).  We also performed aperture photometry
on the mosaics, to determine magnitudes for the objects that lie in
the Wide Field (WF) images, and determined the zeropoint shifts
between WF and PC photometry.  

\begin{figure*}
%\rotate
\figurenum{12}
%% m30_finders_300_rgb_2.eps
\plotone{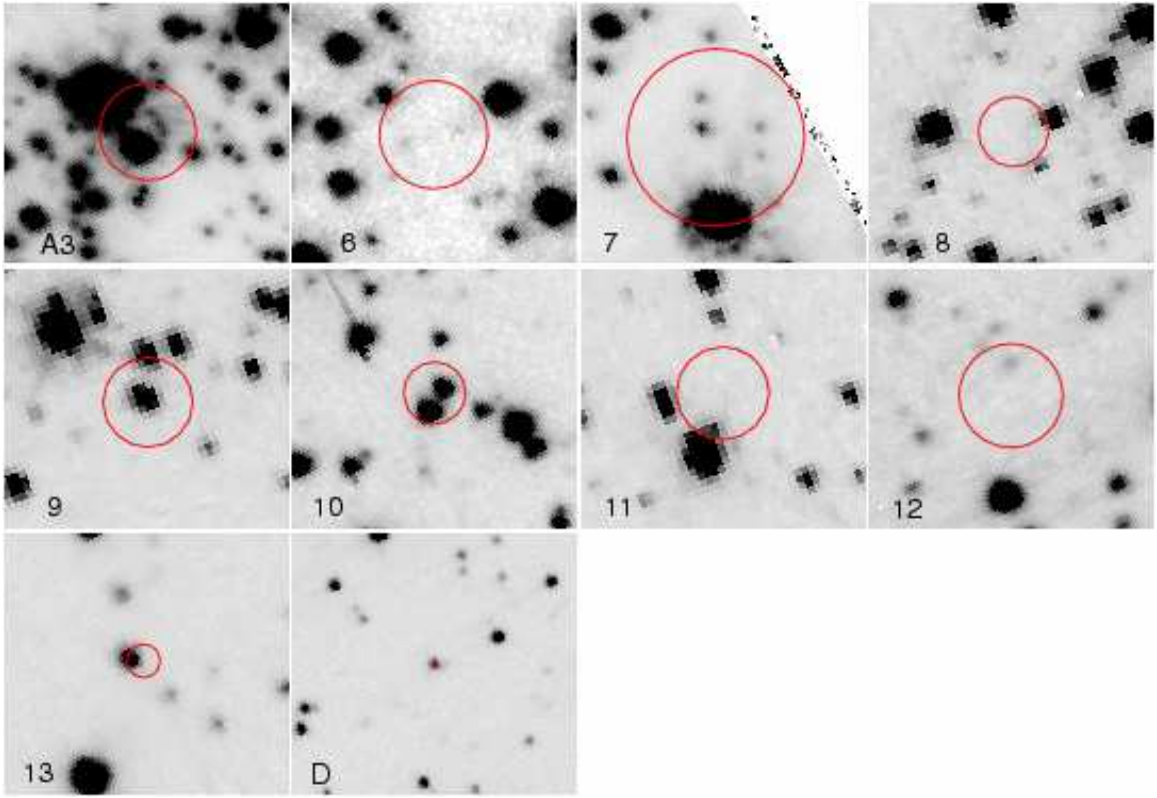}
\figcaption{$V$-band finding charts for remaining X-ray sources within
  the half-mass radius and source~D, which defines the boresight
  correction. North is up and east is to the left.  The east-west
  dimensions are as follows: 2\arcsec\ (sources A3, 6, 7); 4\arcsec\
  (sources 8--13); 8\arcsec\ (source~D).  The error circles represent
  4$\sigma$ uncertainties in the boresight-corrected source positions.
  \emph{(See the electronic edition of the Journal for a color version
  of this figure.)}
\label{fig:v_finders}}
\end{figure*}

Table~\ref{tab:sources} lists the optical counterparts that were
found.  The sources A1, A2, B, and C, 7, and 10 all have possible
counterparts based on positional agreement and optical photometry.
The photometry for the counterparts is summarized in
Table~\ref{tab:counterparts} and is plotted in Fig.~\ref{fig:pc_cmd}.
The positions of the A1, B, C, 7, and 10 counterparts agree to within
about 0\farcs05 with the boresight-corrected \chandra\ positions.  In
the following, we discuss the nature of each of these counterparts as
well as likely counterparts to source D and a few additional sources.
Figure~\ref{fig:u_pc_stamps} shows $U$- and $V$-band finding charts,
from the drizzle-reconstructed images, for the candidate counterparts
to sources A1, A2, B, and C\@.  Figure~\ref{fig:v_finders} shows
$V$-band finding charts for the remaining sources within $r_h$ and
source~D.  

\subsection{Source A1} 

Examination of the drizzle-reconstructed images indicates that there
is a pair of stars with a separation of only about 0\farcs03 within
0\farcs05 of the position of source A1, the likely qLMXB\@.  The
combined photometry for the two stars falls on the main sequence in
the ($V$, \uv) CMD and to the blue side of the MSTO region in the
($V$, \vi) CMD, as shown in Fig.~\ref{fig:pc_cmd}.  The combined
magnitude of the two stars is $V=18.9$.  Since the fainter
(easternmost) of the two stars appears somewhat bluer than the
brighter star (see Fig.~\ref{fig:u_pc_stamps}), it must lie somewhat
to the blue side of the MSTO, making it more likely to be the
counterpart than the brighter, redder star.  We estimate that this
potential counterpart has $V \gtrsim 20$, corresponding to $M_V
\gtrsim 5$.  By comparison, \citet{Edmonds02} have identified an
optical counterpart to the qLMXB X5 in 47~Tuc with $M_V = 8.1$, noting
that two other identified qLMXB counterparts have similar absolute
magnitudes of $M_V \sim 8$.  Given that the candidate counterpart may
be 3 mag brighter than this, it is quite possible that this represents
a case of random positional coincidence in the extremely crowded core
region and that neither of the two stars in the source A1 error circle
is the true optical counterpart to the qLMXB\@.

\subsection{Source A2\label{source_A2}}

As noted in \S\ref{source_detection}, A2 is only detected by WAVDETECT
in a high-energy band (2.5--7~keV), but is detected by PWDETECT in
broad-band images.  In order to better isolate A2 from its brighter,
softer neighbor A1, we used the CIAO DMSTAT task to find the centroid
of the events within a 1\arcsec\ radius circular aperture about the
apparent A2 source location using a range of low-energy cutoff values;
a uniform upper cutoff of 6 keV was adopted.  As the low-energy cutoff
was increased from 0.5 to 2.0 keV, the centroid shifted by 0\farcs2.
The formal error in the centroid is about 0\farcs07 in each
coordinate.  The range of centroided positions for A2 is offset by
0\farcs1 -- 0\farcs3 from an extremely blue star, with the smaller
offset corresponding to the low-energy cutoff values in excess of 1.75
keV.  The star is denoted as \#3327 in the photometry of
\citet{Guhathakurta98}; they find $V=17.84$, $\bv=-0.17$, and
$\uv=-1.07$.  Our aperture photometry similarly gives $V=17.77$,
$\uv=-1.19$, and $\vi=0.02$.  In the ($V$, \vi) CMD, this star is as
blue as the blue tip of the horizontal branch, which represents the
bluest color possible for a black body.  Its position in this CMD is
consistent with the blue straggler region.  However, its \uv\ color is
over 1 mag bluer than the blue stragglers, which lie vertically above
the MSTO in the ($V$, \uv) CMD\@.  Not surprisingly, this star is the
brightest object in the cluster core in \hst\ WFPC2 imaging in the
F160W ultraviolet filter.  Given the extreme colors of this star and
its possible positional consistency with A2, further investigation of
it as a possible counterpart is appropriate.

The optical color indices of the possible A2 counterpart are
consistent with an early B spectral type ($\sim$ B2) and its absolute
magnitude of $M_V = 3.1$ is within the observed range for subdwarf B
(sdB) stars \citep{Edelmann03}.  Common envelope evolution in a close
binary system is a likely scenario for the formation of these extreme
horizontal branch (EHB) and post-EHB stars \citep{Han02,Han03}.  This
is supported by the observation that a substantial fraction
($69\pm9\%$) of sdB stars appear to be in close binary systems
\citep{Maxted01}.  The shortest observed orbital period to date is
1.8~h \citep{Maxted02}.  The binary companion of the sdB star is
likely to be a white dwarf in many cases
\citep[e.g.][]{Napiwotzki01,Maxted02}.  X-ray emission has been
observed from some subdwarf binaries \citep[e.g.][]{Israel97},
although this does not appear to be a common type of source.  Thus, it
appears likely that the near positional coincidence between A2 and the
sdB star represents a chance superposition.  We note that no object
with similar optical properties is observed as a counterpart to a
\chandra\ source in either of the two well-studied clusters 47~Tuc
\citep{Edmonds03} and NGC~6397 \citep{Taylor05}.  Nevertheless,
\chandra\ source A2 is clearly a cluster member given its location
within 1\arcsec\ of the cluster center.  As noted in
\S\ref{source_properties}, it has X-ray color indices consistent with
the likely CVs, sources B and C, which suggests that source A2 is also
likely to be a CV with a faint optical counterpart.  The extremely
crowded region about the position of A2 might well preclude the
optical detection of such a CV, even with relatively deep \hst\
imaging.

\subsection{Source B}

Source B is closely aligned with a faint object that is visible in the
GO-7379 $U$ frame.  In the $V$ and $I$ frames, it lies in the halo of
a giant-branch neighbor, at a separation of just 0\farcs35.  In order
to better estimate its $V$ and $I$ magnitudes, we carried out PSF
subtraction using DAOPHOT, defining the PSF from nearby stars of
comparable brightness to the neighbor.  Given the $\Delta V \approx 7$
difference between the source B counterpart and its neighbor, this
results in large uncertainties in the magnitudes; we estimate these as
$0.2$ mag for $U$ and $I$ and $0.4$ mag in $V$\@.  The resulting
magnitudes and colors are $V = 23.5$ ($M_V = 8.6$), $\uv = -1.0$, and
$\vi = 1.7$.  We have plotted the corresponding location of the source
B counterpart, with error bars, in Fig.~\ref{fig:pc_cmd}.  While the
\vi\ color is consistent with a faint main sequence location, there is
a very large apparent UV excess relative to an extension of lower main
sequence.  These photometric properties are consistent with a CV
interpretation, in which the $V$ and $I$ bands are strongly dominated
by radiation from the secondary, while the $U$ band is dominated by
contributions from the much hotter white dwarf and accretion disk
\citep[c.f.][]{Edmonds03}.  Aperture photometry of the combined $U$
frames for each of the 12 dither positions suggests substantial time
variability.  Based on these optical properties of its apparent
counterpart, its location within 5\arcsec\ of the cluster center, and
the consistency of the X-ray colors of sources B and C, it appears
likely that source B is a CV located near the core of M30.

\subsection{Source C \label{source_C}}

Source C is closely aligned with an object located 12\arcsec\ from the
cluster center with $V = 20.94$ ($M_V = 6.1$), $\uv = -0.44$, and
$\vi=0.89$.  While the \vi\ color is just to the red of the main
sequence, the \uv\ color is 0.6 mag to the blue of the main sequence
(see Fig.~\ref{fig:pc_cmd}).  As for source B, this is entirely
consistent with a CV interpretation.  Curiously, this counterpart does
not appear in the \citet{Guhathakurta98} photometry, although other
nearby stars of similar magnitude are detected.  A check of the
GO-5324 F336W image, on which the \citet{Guhathakurta98} photometry is
based, indicates that this candidate counterpart to source C is
present, but is located within 3 pixels of the vignetted region of the
PC field.  This may account for it being excluded from the final
photometry list.  This object is well detected in the single 600\,s
F336W exposure from GO-5603 (obtained in 1994).  Approximate aperture
photometry indicates that the object is about 1.0 mag fainter in
F336W, in 1999 versus 1994. This high level of variability provides
additional support for the CV interpretation.

\subsection{Source D\label{source_D}}

As previously mentioned, our boresight correction is defined by source
D\@. The likely counterpart has $V = 20.74$, a \vi\ color consistent
with the main sequence and a \uv\ color that is 0.6 mag to the blue of
the main sequence.  Thus, its photometric properties appear quite
similar to those of source C, as can be seen in
Fig.~\ref{fig:pc_cmd}. This argues for interpreting source D as a CV
in M30, which would be quite interesting given its location at
$1.4~r_h$.  However, the inferred $L_X = 2\times10^{33}~\ergs$ for
source D would put it among the most luminous CVs seen in any globular
cluster.  In addition, its \chandra\ spectrum, discussed in
\S\ref{other_sources} above, is not well fit by a MEKAL spectrum,
unlike those of sources A2, B, and C.

In order to obtain further information on the source D counterpart, we
obtained optical spectra using the Hydra spectrographs on the CTIO
4\,m telescope in 2003 July and on the WIYN 3.5\,m telescope in 2003
September.  The seeing was poor throughout the CTIO run, typically
exceeding 2\farcs5.  Thus, the light of this object was strongly mixed
with that of its much brighter neighbor at 2\arcsec\ separation and
only a combined spectrum was obtained.  While the seeing at WIYN was
generally better, the large airmass of this southern cluster resulted
in a similar mixing of the star with its neighbor.  Examination of
this combined spectrum indicates a emission line at 5562~\AA\ with a
FWHM of $\sim12$~\AA, with no evidence for Balmer emission at the
cluster velocity.  While the detection of only one emission line
precludes a definitive redshift determination, we note that
identifying this as the relatively isolated \ion{Mg}{2} $\lambda2800$
line would imply a redshift of $z=0.99$.  A long-slit spectrum of the
optical counterpart to source D would greatly assist in determining
its physical nature.

\subsection{Other Identifications\label{other_IDs}}

Two additional sources visible in Fig.~\ref{fig:acis_64_all}, source 6
(CXOGLB J214021.48-231054.5) and source 7 (CXOGLB
J214021.56-231032.6), fall within the drizzled Planetary Camera field.
No object is detected in the $U$-band image within the error circle
for source 6, while there are two very faint stars within the error
circles in the $V$- and $I$-band images (see
Fig.~\ref{fig:v_finders}).  Source 7 is very close to the edge of the
PC field and was imaged in 8 of 12 pointings.  Three faint objects are
detected within the error circles in all three bands for source 7 and
one additional faint object is detected in the $V$ and $I$ bands (see
Fig.~\ref{fig:v_finders}).  None of these objects appears to be blue.
Photometry for the brightest of the three is shown in
Fig.~\ref{fig:pc_cmd}.  While this object lies on the main sequence in
the ($V$, \vi) CMD, it has an apparent UV deficit and is thus not
likely to be a CV\@.  We note that active binary (AB) counterparts to
\chandra\ sources in globular clusters typically lie on the main
sequence or somewhat above it, as seen in the clusters NGC~6397
\citep{Grindlay01b}, NGC~6752 \citep{Pooley02a}, 47~Tuc
\citep{Edmonds03}, and M4 \citep{Bassa04}.  Thus, the CMD location of
this possible counterpart to source 7 is consistent with an active
binary interpretation.

It can be seen in Fig.~\ref{fig:v_finders} that optical objects are
detected in several of the source error circles in the WF fields.  The
source 9 error circle contains a close pair of objects.  This location
has only been imaged by the GO-5630 program, which provides F555W
($V$) and F160BW coverage.  The images in the latter band are rather
shallow.  Since the possible source 9 counterparts are only detected
in F555W, no color information is available.  Of the two objects in
the source 10 error circle, the southern object lies on the main
sequence in both CMDs, while the northern object lies above the main
sequence in both CMDs (see Fig.~\ref{fig:pc_cmd}).  Thus, this latter
object is consistent with being an active binary.  The faint object
detected near the edge of the source 12 error circle in the $V$ and
$I$ frames is not detected in the $U$ frame.  It is thus likely to be
a faint main sequence star.  The object detected on the edge of the
error circle for source 13 lies on the main sequence in both CMDs.
Thus, this object does not appear likely to be the source 13
counterpart, despite its close proximity to the source 13 location.

We attempted to determine optical counterparts for additional
\chandra\ sources listed in Table~\ref{tab:sources}, by comparison of
the \chandra, \hst, and ground-based images, and then obtaining Hydra
spectroscopy for candidate counterparts.  As noted above, the crowded
conditions, even outside of the half-mass radius, proved challenging
for the spectroscopic program.  At both CTIO and WIYN, the minimum
fiber size is 2\arcsec.  The goal of this effort was to determine
whether any other \chandra\ sources are likely to be associated with
cluster objects, particularly for sources, such as D, that lie beyond
the half-mass radius.  Any source in the cluster halo that is
confirmed as a cluster member is a good candidate for having been
ejected from the cluster core by an energetic binary interaction.

Chandra source 39 (CXOU J214033.30-231236.3) at 3\farcm2 from the
cluster center coincides with a bright star that is almost certainly a
foreground object.  This star is listed as \#1 in the photometry of
\citet{Sandquist99} and has $V = 12.03, \vi = 0.80$, which place it at
the magnitude level of the tip of the giant branch but 0.6 mag bluer
in \vi\ color.  The color index indicates a spectral type of
approximately G2\@. The X-ray to optical flux ratio of $F_X/F_V
\approx 10^{-4}$ is within the range observed for main-sequence early
G stars \citep{Huensch98}.  The soft X-ray spectrum of this source
provides additional evidence that it is a foreground star (see
\S\ref{other_sources}).

The CTIO and WIYN Hydra spectroscopy described previously resulted in
the tentative identification of three of the other Chandra sources,
numbers 16, 27, and 50 in Table~\ref{tab:sources} (CXOU
J214016.05-230957.3, 214015.01-231217.5, and 214037.03-231337.1,
respectively).  These objects are located at distances of 1\farcm6,
2\farcm2, and 4\farcm4 respectively, from the cluster center.  While
the spectra of the apparent counterparts to sources 16 and 27 are
blended with neighboring stars, there are broad emission lines
apparent in each of the spectra that give consistent redshifts.  These
lines are enhanced when a smooth continuum is fit and subtracted.  For
source 16, Ly-$\alpha$, \ion{C}{4} $\lambda$1549, and \ion{C}{3}]
$\lambda$1909 give a mean redshift of $2.42$.  For source 27,
Ly-$\alpha$, \ion{N}{5} $\lambda$1240, \ion{Si}{4}+\ion{O}{4}]
$\lambda$ 1400, and \ion{C}{4} $\lambda$1549 give a mean redshift of
$3.08$.  Source 50 is positionally coincident with an apparent
elliptical galaxy with a redshift of 0.27, based on a detection of the
Mg~b triplet and the \ion{Ca}{2} H \& K break. It is not surprising
that these three sources, which all lie well outside of the cluster
half-mass radius, are background objects.  However, this does not
resolve the issue of whether a small number of the other sources near
or outside of $r_h$ may actually be cluster members that were ejected
from the central region.  This will require additional spectroscopy,
under more favorable observing conditions.  The use of a long slit or
slitlets would facilitate separation of objects from close neighbors.

\begin{figure}
\figurenum{13}
%% star091.eps
\plotone{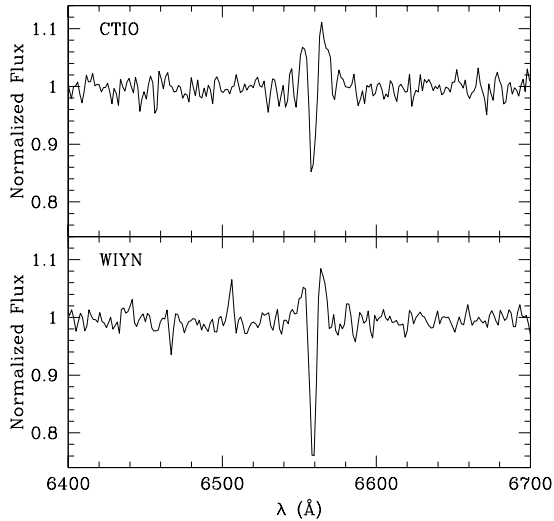}
\figcaption{\ha\ line profiles from Hydra multiobject spectroscopy of
  possible counterpart to \chandra\ source 36.  The CTIO spectrum
  (top) was obtained in 2003 July; the WIYN spectrum (bottom) was
  obtained in 2003 September.  The flux has been normalized to the
  continuum in both cases.  Note the apparent blue and red emission
  wings about \ha, which suggest the superposition of a broad emission
  feature on a narrower absorption line.  The apparent weak emission
  feature near 6505~\AA\ in the WIYN spectrum is likely noise, given
  its absence in the CTIO spectrum.
\label{fig:ha_emission}}
\end{figure}

In addition to the tentative identifications just described, source 40
(CXOU J214008.37-231118.9) lies within about 1\arcsec\ of a $V=16.4$
star, located at 3\farcm2 from the cluster center, that lies about
0.06 mag to the blue of the ridgeline of the giant branch in the
($V$, \vi) CMD of \citet{Sandquist99}.  This star is also detected by
\citet{Richer88} with nearly the same $V$ magnitude ($V=16.5$) and a
\bv\ color that is about 0.08 mag to the blue of the giant branch in
the ($V$, \bv) CMD\@.  Examination of the object in deep $B$, $V$, and
$I$ CTIO 1.5\,m images (kindly provided by M.~Bolte) indicates an
elongation to the northeast that becomes more prominent at shorter
wavelength.  This suggests that the optical object may be the
superposition of a faint blue star with a much brighter giant.  The
separation between the two apparent objects is of order 0\farcs5.  The
spectra for the combined object, obtained at both CTIO and WIYN, show
definite emission components in the red and blue wings of the \ha\
absorption line, as can be seen in Figure~\ref{fig:ha_emission}.  This
suggests that the putative blue object has a broad single-component
\ha\ emission feature that overlies the absorption line of the giant.
Given the small offset between this apparent emission object and the
much brighter giant, further spectroscopic study will require high
angular resolution.  If a broad \ha\ emission feature at a velocity
consistent with cluster membership were confirmed, then source 40 may
be consistent with being a CV that was ejected from the cluster core.

We checked the positions of the eight variable stars that were
identified by \citet{Pietrukowicz04} within the GO-7379 field.  We
find that none of these stars is positionally coincident with a
\chandra\ source.

\section{Discussion}

It is interesting to ask whether the observed size of the X-ray source
population in M30 accords with expectation.  \citet{Pooley03} have
shown that for 12 globular clusters observed by \chandra, the X-ray
source population size scales approximately linearly with the
predicted stellar encounter rate.  They interpret this as strong
empirical evidence that the long-observed overabundance of exotic
binaries in clusters is primarily the result of dynamical formation in
the dense core regions.  Following \citet{Verbunt87}, they define a
normalized encounter rate $\Gamma$ as the volume integral to the
half-mass radius of $\rho^2/v$, where $\rho$ is the total mass
density, $v$ is the velocity dispersion, and single-mass
\citet{King66} model structure is adopted for the radial profiles of
$\rho$ and $v$.  Most of the contribution to $\Gamma$ comes from the
core and the immediately surrounding region; a simple computation
using an analytic King model shows that about 90\% of the interactions
occur within $2 r_c$.  As noted by \citet{Pooley03}, this formalism
leads to the scaling $\Gamma \propto \rho_0^2 \, r_c^3/v_0$.

M30 is one of the clusters included in the \citet{Pooley03} sample and
their estimate of its background-corrected population size of
$\sim$5--6 \chandra\ sources within the half-mass radius agrees well
with their prediction of $\sim$5 sources from their empirical
$N$--$\Gamma$ relation.  Our estimate of the background-corrected
source population size is twice as large, at $10\pm2$, where the error
represents the uncertainty in the background correction only, as do
the error bars in Fig.~2 of \citet{Pooley03}. Our background-corrected
source count suggests that M30 has a source excess relative to its
computed interaction rate $\Gamma$, although small number statistics
are clearly an issue here.

Somewhat surprisingly, M30 has the fourth smallest interaction rate in
the \citet{Pooley03} sample, notwithstanding its very high central
density.  This is a consequence of its small core size.
\citet{Pooley03} adopted values of core radius, central density, and
central velocity dispersion from the \citet{Harris96} compilation
(2003 February online version); the values for M30 are $r_c =
3\farcs6$ and $\rho_0 = 1.1\times10^5~\msun~{\rm pc}^{-3}$.  We note
that given the collapsed-core status of M30, these values are rather
uncertain.  \citet{Sosin97} has placed an upper limit of $1\farcs9$ on
the core radius, suggesting a preferred value of $1\arcsec$, and has
estimated a central density of $8.5\times10^5~\msun~{\rm pc}^{-3}$
from dynamical modeling.  These large changes in $r_c$ and $\rho_0$
produce partially compensating changes in the value of $\Gamma$,
increasing it by a factor of 1.3 for $v_0$ held fixed and by a factor
of 1.7 for $v_0 \propto \rho_0^{0.5} r_c$ (as for a King model).
Another source of uncertainty is that a single-mass King model does
not give a good description of the collapsed-core structure of
M30\@. It is interesting to note that another of the three
core-collapsed clusters in the \citet{Pooley03} sample, NGC~6397,
significantly deviates from the empirical $N$--$\Gamma$ relation in
the sense of having an overabundance of X-ray sources relative to the
encounter rate computed from the \citet{Harris96} values of core
parameters, including a core radius of $3\arcsec$.  We note that even
with an HST-based analysis of the star count profile of NGC~6397, the
core radius value remains uncertain.  \citet{Taylor05} find a best fit
value of $4\farcs4 \pm 3\farcs2$ (1-$\sigma$ errors).  This
uncertainty range leads to a corresponding uncertainty in $\Gamma$,
since $\Gamma \propto r_c$ if the central surface brightness and
central velocity dispersion are fixed.

Given the uncertainties arising from the core-collapsed status of M30
and its consequently small core size, we suggest that its relation to
the other clusters in the \citet{Pooley03} sample is not entirely
clear.  Simple models for core collapse produce the scalings $\rho_0
\propto r_c^{-2.2}$ and $v_0 \propto r_c^{-0.05}$ \citet{Cohn80}.
These result in $\Gamma \propto r_c^{-1.4}$, indicating that clusters
undergoing core radius oscillations should experience episodic bursts
of strongly enhanced binary formation and ejection as the core passes
through its densest phases.  Thus, the size of the X-ray binary
population in a collapsed-core cluster may undergo substantial
fluctuations in time.  In light of these issues, detailed dynamical
simulations of the evolution of binary populations in clusters
undergoing core collapse would be very helpful.

\section{Summary}

Our \chandra\ ACIS-S imaging of the collapsed-core globular cluster
M30 has detected a centrally concentrated population of X-ray sources,
including six within 12\arcsec\ of the cluster center and three within
the 1\farcs9 core radius.  There is a net excess of about 10 sources
above background within the half-mass radius.  The brightest core
source (A1) has an X-ray flux and spectrum that are both consistent
with it being a qLMXB. The other central sources (A2, A3, B, and C)
all have X-ray properties consistent with being CVs.  We have
identified likely optical counterparts to sources B and C that are
similarly consistent with a CV interpretation.  While we have
suggested possible counterparts to sources A1 and A2, the status of
these is uncertain and will require further investigation.  Sources 7
and 10, which lie at 17\arcsec\ and 51\arcsec\ from the cluster center,
respectively, have possible active binary counterparts. There remains a
possibility of a significant number of additional faint sources in and
near the core that lie somewhat below our detection limit of about 4.5
counts (corresponding to to a source luminosity of
$\sim4\times10^{30}~\ergs$).  Additional \chandra\ imaging would be
very useful in searching for a fainter source population.

Our analysis of the radial distribution of the sources suggests that
those belonging to the cluster are largely confined to the half-mass
radius.  There is some evidence for a small excess of sources over the
expected background between $r_h$ and $2 r_h$; some of these may have
been ejected from the central regions.  The radial profile of the
sources within $r_h$ implies a typical source mass of about
$1.3\pm0.2\,\msun$, which is consistent with a variety of binary
systems, including CVs.

Our analysis of the X-ray spectrum of source A1 indicates that it is
best fit by a model for a neutron star thermally radiating through a
hydrogen atmosphere, consistent with the qLMXB interpretation.  While
the spectral modeling does not provide tight constraints on the
neutron star mass and radius, it does demonstrate consistency with
standard neutron star structure models.

Our effort to identify additional sources using ground-based optical
spectroscopy with the CTIO and WIYN Hydra multiobject spectrographs
demonstrated that two of the sources beyond the half-mass radius are
quasars.  The optical spectrum of source D, the brightest source in
the field, shows no evidence for \ha\ emission, while it does show a
possible redshifted emission line.  Thus it is more likely consistent
with a background AGN than with a cluster CV\@.  The spectroscopy did
identify an object, also located beyond the half-mass radius, that
shows \ha\ emission wings about an absorption line.  This object
appears to be the superposition of two stars that are not well
resolved by ground-based imaging and are outside of the field of HST
imaging.  Further investigation of this intriguing object is clearly
warranted.

The core-collapsed status of M30 complicates a comparison of its
\chandra\ source population size $N$ with the empirical relation
between $N$ and collision rate $\Gamma$ established for other
clusters.  While the observed population of about 10 sources above
background within the half-mass radius is twice as large as the
prediction of the relation, the determination of an appropriate
$\Gamma$ value for M30 is challenging given its very small core size
and strong deviation from King-model structure in the inner region.  A
deeper X-ray luminosity function and detailed modeling of source
population evolution in a collapsed-core cluster would facilitate
further investigation of this issue.

\acknowledgements{We are grateful to Mike Bolte and Ron Gilliland for
graciously providing ground-based images of M30 for our study.  We
thank Xavier Koenig and Allen Rogel for obtaining Hydra spectra of
candidate source counterparts.  We also appreciate the helpful
suggestions of the referee.

This project was supported by NASA \chandra\ grant GO2-3037X to
Indiana University.}

%\clearpage

% \bibliography{m30}

\clearpage

\LongTables
\newcommand{\ns}{\hspace*{-2pt}}
\tabletypesize{\small}
\setlength{\tabcolsep}{5.5pt}
\begin{landscape}
\begin{deluxetable}{lccccccccccc}
\tablenum{1}
%\rotate
\tablecolumns{12}
%\tablewidth{0pt}
\tablecaption{\textbf{M30 X-ray Source Properties} \label{tab:sources}}
\tablehead{
\colhead{Source} & 
\colhead{ID} & 
\colhead{Type,} & 
\colhead{RA}  & 
\colhead{Dec} & 
\colhead{$r$} &  
\colhead{Cts} &     
\colhead{$C_1$} & 
\colhead{$C_2$} &
\colhead{$F_X$} &
\colhead{$F_X$} &  
\colhead{Name}  \\ 
 &  & [redshift] & (h m s err) & (\degr\ \arcmin\ \arcsec\ err)
 &  (\arcmin) & \ns(0.3--7\,keV)\ns &  &
 &  \ns(0.5--6\,keV)\ns  &  \ns(0.5--2.5\,keV)\ns  &  (positional part) \\  
(1) & (2) & (3) & (4) & (5) & (6) & (7) & (8) & (9) & (10) & (11) & (12) 
} 
\startdata
\cutinhead{\textbf{Sources within \boldmath $r_h$; designated CXOGLB J\dots}}
1  & A2  &  CV?    &  21:40:22.186  0.003 & $-$23:10:47.20  0.03 & 0.01 & 
   $37.9^{+7.2}_{-6.1}$ & $0.23^{+0.41}_{-0.41}$ & $-0.11^{+0.57}_{-0.58}$ & 
   $6.2^{+3.1}_{-1.4}$ & $3.8^{+1.3}_{-0.8}$ & 214022.18-231047.2 \\[7.5pt]
2  & A1  &  qLMXB  &  21:40:22.130  0.001 & $-$23:10:45.57  0.01 & 0.03 & 
   $900.8^{+31.0}_{-30.0}$ & $-0.95^{+0.08}_{-0.08}$ & $-1.31^{+0.17}_{-0.17}$ & 
   $65.5^{+2.7}_{-2.4}$ & $65.5^{+2.7}_{-2.4}$ & 214022.13-231045.5 \\[7.5pt] 
3  & A3  &         &  21:40:21.995  0.006 & $-$23:10:47.21  0.09 & 0.03 & 
   $9.9^{+4.3}_{-3.1}$ & $0.01^{+0.90}_{-0.90}$ & $-0.44^{+1.38}_{-1.48}$ & 
   $1.0^{+0.6}_{-0.3}$ & $1.0^{+0.6}_{-0.3}$ & 214021.99-231047.2 \\[7.5pt] 
4  & B  & CV       &  21:40:22.153  0.002 & $-$23:10:51.66  0.04 & 0.07 & 
   $97.6^{+10.9}_{-9.9}$ & $0.74^{+0.26}_{-0.26}$ & $-0.04^{+0.30}_{-0.30}$ & 
   $12.7^{+2.9}_{-1.8}$ & $7.0^{+1.1}_{-0.8}$ & 214022.15-231051.6 \\[7.5pt] 
5  & C  & CV       &  21:40:22.929  0.002 & $-$23:10:49.25  0.03 & 0.19 & 
   $122.5^{+12.1}_{-11.1}$ & $0.96^{+0.24}_{-0.24}$ & $0.87^{+0.28}_{-0.28}$ & 
   $20.1^{+3.4}_{-2.4}$ & $7.9^{+1.2}_{-0.9}$ & 214022.92-231049.2 \\[7.5pt] 
6  &   &           &  21:40:21.482  0.011 & $-$23:10:54.57  0.06 & 0.19 & 
   $7.6^{+3.9}_{-2.7}$ & $-0.54^{+1.06}_{-1.12}$ & $>-0.15$ &
   $1.0^{+0.9}_{-0.4}$ & $0.5^{+0.6}_{-0.2}$ & 214021.48-231054.5 \\[7.5pt] 
7  &   & AB?       &  21:40:21.564  0.015 & $-$23:10:32.60  0.12 & 0.28 & 
   $4.6^{+3.3}_{-2.1}$ & $-0.44^{+1.41}_{-1.50}$ & $>0.50$ & 
   $0.3^{+0.3}_{-0.1}$ & $0.3^{+0.3}_{-0.1}$ & 214021.56-231032.6 \\[7.5pt] 
8  &   &           &  21:40:22.109  0.011 & $-$23:11:14.10  0.10 & 0.45 & 
   $11.7^{+4.5}_{-3.4}$ & $0.75^{+0.89}_{-0.84}$ & $1.19^{+1.33}_{-1.16}$ & 
   $2.0^{+2.0}_{-0.7}$ & $0.9^{+0.5}_{-0.3}$ & 214022.10-231114.1 \\[7.5pt] 
9  &   &           &  21:40:20.363  0.014 & $-$23:10:22.88  0.13 & 0.57 & 
   $6.6^{+3.7}_{-2.5}$ & $-0.97^{+1.21}_{-1.37}$ & $<0.47$ & 
   $0.4^{+0.3}_{-0.2}$ & $0.4^{+0.3}_{-0.2}$ & 214020.36-231022.8 \\[7.5pt] 
10  &  & AB?       &  21:40:23.216  0.007 & $-$23:09:58.49  0.12 & 0.85 & 
   $6.5^{+3.7}_{-2.5}$ & $0.34^{+1.19}_{-1.15}$ & $-0.01^{+1.64}_{-1.64}$ & 
   $0.6^{+0.7}_{-0.3}$ & $0.3^{+0.3}_{-0.1}$ & 214023.21-230958.4 \\[7.5pt] 
11  &   &          &  21:40:19.361  0.012 & $-$23:11:24.38  0.16 & 0.89 & 
   $5.6^{+3.5}_{-2.3}$ & $>-1.08$ & $0.74^{+1.41}_{-1.28}$ & 
   $1.0^{+0.9}_{-0.4}$ & $0.4^{+0.6}_{-0.2}$ & 214019.36-231124.3 \\[7.5pt] 
12  &   &          &  21:40:26.981  0.015 & $-$23:10:38.84  0.16 & 1.13 & 
   $5.6^{+3.5}_{-2.3}$ & $1.81^{+1.92}_{-1.63}$ & $1.51^{+1.96}_{-1.66}$ & 
   $0.9^{+1.1}_{-0.4}$ & $0.3^{+0.3}_{-0.1}$ & 214026.98-231038.8 \\[7.5pt] 
13  &   &          &  21:40:26.546  0.005 & $-$23:11:18.26  0.05 & 1.14 & 
   $14.6^{+4.9}_{-3.8}$ & $2.94^{+1.82}_{-1.47}$ & $2.81^{+1.83}_{-1.45}$ & 
   $6.0^{+2.8}_{-1.8}$ & $0.2^{+0.3}_{-0.1}$ & 214026.54-231118.2 \\[7.5pt] 
\cutinhead{\textbf{Sources outside of \boldmath $r_h$; designated CXOU J\dots}}
14  &   &          &  21:40:17.349  0.011 & $-$23:11:34.13  0.08 & 1.35 & 
   $6.5^{+3.7}_{-2.5}$ & $0.99^{+1.38}_{-1.22}$ & $0.43^{+1.50}_{-1.41}$ & 
   $1.1^{+1.2}_{-0.4}$ & $0.5^{+0.6}_{-0.2}$ & 214017.34-231134.1 \\[7.5pt] 
15  &   &          &  21:40:23.831  0.015 & $-$23:09:20.30  0.15 & 1.50 & 
   $5.5^{+3.5}_{-2.3}$ & $1.76^{+1.92}_{-1.64}$ & $>-0.79$ & 
   $1.6^{+2.0}_{-0.7}$ & $0.4^{+0.6}_{-0.2}$ & 214023.83-230920.3 \\[7.5pt] 
16  &   &  QSO, 2.43  &  21:40:16.055  0.004 & $-$23:09:57.36  0.03 & 1.63 & 
   $98.4^{+11.0}_{-9.9}$ & $-0.20^{+0.24}_{-0.24}$ & $-0.56^{+0.38}_{-0.39}$ & 
   $8.6^{+2.3}_{-1.1}$ & $6.9^{+1.1}_{-0.8}$ & 214016.05-230957.3 \\[7.5pt]
17  & D  &  QSO?   &  21:40:29.235  0.001 & $-$23:10:31.30  0.01 & 1.65 & 
   $1850.7^{+44.0}_{-43.0}$ & $0.25^{+0.05}_{-0.05}$ & $0.22^{+0.07}_{-0.07}$ & 
   $195.2^{+7.4}_{-6.6}$ & $111.2^{+3.4}_{-3.2}$ & 214029.23-231031.3 \\[7.5pt] 
18  &   &          &  21:40:22.946  0.017 & $-$23:09:06.09  0.10 & 1.70 & 
   $4.3^{+3.2}_{-2.0}$ & $>-0.77$ & $>-0.77$ & 
   $1.3^{+1.9}_{-0.6}$ & $0.3^{+0.6}_{-0.2}$ & 214022.94-230906.0 \\[7.5pt] 
19  &   &          &  21:40:20.815  0.003 & $-$23:12:29.93  0.07 & 1.74 & 
   $33.7^{+6.9}_{-5.8}$ & $1.91^{+0.68}_{-0.63}$ & $1.04^{+0.55}_{-0.53}$ & 
   $6.4^{+2.7}_{-1.5}$ & $2.6^{+0.9}_{-0.6}$ & 214020.81-231229.9 \\[7.5pt] 
20  &   &          &  21:40:28.337  0.011 & $-$23:09:29.96  0.17 & 1.92 & 
   $13.5^{+4.8}_{-3.6}$ & $1.41^{+1.00}_{-0.90}$ & $1.61^{+1.28}_{-1.09}$ & 
   $2.6^{+2.1}_{-1.0}$ & $1.0^{+0.7}_{-0.3}$ & 214028.33-230929.9 \\[7.5pt] 
21  &   &          &  21:40:18.345  0.005 & $-$23:12:32.20  0.07 & 1.95 & 
   $32.6^{+6.8}_{-5.7}$ & $1.24^{+0.54}_{-0.52}$ & $0.43^{+0.54}_{-0.53}$ & 
   $4.0^{+1.6}_{-0.8}$ & $2.4^{+0.8}_{-0.5}$ & 214018.34-231232.2 \\[7.5pt] 
22  &   &          &  21:40:29.878  0.008 & $-$23:09:57.06  0.23 & 1.97 & 
   $7.3^{+3.8}_{-2.6}$ & $1.27^{+1.41}_{-1.21}$ & $>-1.07$ & 
   $2.3^{+2.0}_{-1.0}$ & $0.2^{+0.3}_{-0.1}$ & 214029.87-230957.0 \\[7.5pt] 
23  &   &          &  21:40:14.055  0.018 & $-$23:11:42.50  0.10 & 2.07 & 
   $5.5^{+3.5}_{-2.3}$ & $-0.78^{+1.29}_{-1.43}$ & $0.00^{+2.42}_{-2.42}$ & 
   $0.6^{+0.7}_{-0.3}$ & $0.3^{+0.3}_{-0.1}$ & 214014.05-231142.5 \\[7.5pt] 
24  &   &          &  21:40:30.959  0.016 & $-$23:10:18.78  0.13 & 2.08 & 
   $11.4^{+4.5}_{-3.3}$ & $>-1.93$ & $>-1.93$ & 
   $4.7^{+2.6}_{-1.5}$ & $0.4^{+0.6}_{-0.3}$ & 214030.95-231018.7 \\[7.5pt] 
25  &   &          &  21:40:18.087  0.012 & $-$23:12:40.51  0.12 & 2.11 & 
   $6.6^{+3.7}_{-2.5}$ & $1.05^{+1.41}_{-1.23}$ & $-0.45^{+1.39}_{-1.49}$ & 
   $0.7^{+0.9}_{-0.3}$ & $0.3^{+0.3}_{-0.1}$ & 214018.08-231240.5 \\[7.5pt] 
26  &   &          &  21:40:15.717  0.003 & $-$23:09:11.73  0.03 & 2.17 & 
   $200.8^{+15.2}_{-14.2}$ & $0.69^{+0.17}_{-0.17}$ & $0.50^{+0.21}_{-0.21}$ & 
   $23.3^{+3.1}_{-2.1}$ & $14.5^{+1.5}_{-1.3}$ & 214015.71-230911.7 \\[7.5pt] 
27  &   & QSO, 3.08     &  21:40:15.015  0.004 & $-$23:12:17.56  0.04 & 2.22 & 
   $83.5^{+10.2}_{-9.1}$ & $0.52^{+0.27}_{-0.27}$ & $-0.01^{+0.34}_{-0.34}$ & 
   $9.5^{+2.5}_{-1.4}$ & $6.1^{+1.1}_{-0.8}$ & 214015.01-231217.5 \\[7.5pt]
28  &   &          &  21:40:32.027  0.009 & $-$23:11:46.94  0.29 & 2.48 & 
   $5.6^{+3.5}_{-2.3}$ & $>-1.09$ & $>-1.09$ & 
   $1.3^{+1.0}_{-0.5}$ & $0.8^{+0.7}_{-0.4}$ & 214032.02-231146.9 \\[7.5pt] 
29  &   &          &  21:40:12.567  0.015 & $-$23:09:36.80  0.12 & 2.49 & 
   $19.0^{+5.4}_{-4.3}$ & $2.47^{+1.30}_{-1.06}$ & $2.28^{+1.26}_{-1.04}$ & 
   $4.3^{+2.5}_{-1.2}$ & $1.4^{+0.8}_{-0.4}$ & 214012.56-230936.8 \\[7.5pt] 
30  &   &          &  21:40:33.237  0.006 & $-$23:10:21.26  0.10 & 2.59 & 
   $44.2^{+7.7}_{-6.6}$ & $2.06^{+0.61}_{-0.57}$ & $0.87^{+0.45}_{-0.44}$ & 
   $7.6^{+2.0}_{-1.4}$ & $2.8^{+0.9}_{-0.6}$ & 214033.23-231021.2 \\[7.5pt] 
31  &   &          &  21:40:31.449  0.009 & $-$23:09:15.06  0.09 & 2.63 & 
   $28.1^{+6.4}_{-5.3}$ & $0.37^{+0.49}_{-0.49}$ & $0.64^{+0.71}_{-0.68}$ & 
   $4.5^{+2.3}_{-1.1}$ & $1.9^{+0.8}_{-0.4}$ & 214031.44-230915.0 \\[7.5pt] 
32  &   &          &  21:40:20.941  0.009 & $-$23:08:01.43  0.07 & 2.78 & 
   $91.6^{+10.6}_{-9.6}$ & $0.16^{+0.25}_{-0.25}$ & $0.42^{+0.36}_{-0.36}$ & 
   $10.7^{+2.6}_{-1.5}$ & $6.4^{+1.1}_{-0.8}$ & 214020.94-230801.4 \\[7.5pt] 
33  &   &          &  21:40:16.192  0.004 & $-$23:08:19.79  0.03 & 2.81 & 
   $357.6^{+19.9}_{-18.9}$ & $0.10^{+0.12}_{-0.12}$ & $0.62^{+0.18}_{-0.18}$ & 
   $39.9^{+3.9}_{-2.9}$ & $23.8^{+1.8}_{-1.6}$ & 214016.19-230819.7 \\[7.5pt] 
34  &   &          &  21:40:21.699  0.008 & $-$23:13:36.09  0.17 & 2.82 & 
   $11.8^{+4.5}_{-3.4}$ & $1.19^{+1.01}_{-0.92}$ & $1.36^{+1.31}_{-1.13}$ & 
   $4.7^{+4.4}_{-1.7}$ & $1.2^{+0.8}_{-0.4}$ & 214021.69-231336.0 \\[7.5pt] 
35  &   &          &  21:40:09.644  0.021 & $-$23:10:46.20  0.17 & 2.87 & 
   $5.9^{+3.6}_{-2.4}$ & $>-1.23$ & $>-1.23$ & 
   $2.9^{+2.4}_{-1.1}$ & $0.2^{+0.5}_{-0.2}$ & 214009.64-231046.2 \\[7.5pt] 
36  &   &          &  21:40:34.509  0.012 & $-$23:09:47.00  0.25 & 3.02 & 
   $9.1^{+4.1}_{-2.9}$ & $2.52^{+1.74}_{-1.58}$ & $2.28^{+1.87}_{-1.53}$ & 
   $2.2^{+1.9}_{-0.7}$ & $1.0^{+0.8}_{-0.4}$ & 214034.50-230947.0 \\[7.5pt] 
37  &   &          &  21:40:30.865  0.005 & $-$23:13:08.23  0.10 & 3.09 & 
   $38.8^{+7.3}_{-6.2}$ & $3.19^{+1.20}_{-0.98}$ & $0.93^{+0.47}_{-0.45}$ & 
   $19.9^{+6.4}_{-3.9}$ & $4.9^{+2.1}_{-1.1}$ & 214030.86-231308.2 \\[7.5pt] 
38  &   &          &  21:40:32.697  0.014 & $-$23:08:48.88  0.27 & 3.13 & 
   $9.9^{+4.3}_{-3.1}$ & $0.23^{+0.91}_{-0.89}$ & $>-1.01$ & 
   $2.0^{+2.0}_{-0.7}$ & $0.8^{+0.7}_{-0.3}$ & 214032.69-230848.8 \\[7.5pt] 
39  &   &  star\tablenotemark{a}   &  21:40:33.308  0.004 & $-$23:12:36.35  0.09 & 3.15 & 
   $56.5^{+8.6}_{-7.5}$ & $-1.22^{+0.38}_{-0.39}$ & $-1.01^{+0.81}_{-0.86}$ & 
   $5.2^{+1.1}_{-0.7}$ & $5.2^{+1.1}_{-0.7}$ & 214033.30-231236.3 \\[7.5pt] 
40  &   &          &  21:40:08.371  0.020 & $-$23:11:18.99  0.13 & 3.20 & 
   $10.6^{+4.4}_{-3.2}$ & $1.92^{+1.45}_{-1.19}$ & $0.38^{+0.97}_{-0.94}$ & 
   $1.8^{+2.0}_{-0.7}$ & $0.9^{+0.7}_{-0.3}$ & 214008.37-231118.9 \\[7.5pt] 
41  &   &          &  21:40:08.077  0.028 & $-$23:11:30.69  0.16 & 3.31 & 
   $9.2^{+4.2}_{-3.0}$ & $1.65^{+1.37}_{-1.15}$ & $>-1.49$ & 
   $2.4^{+2.0}_{-0.9}$ & $0.4^{+0.4}_{-0.2}$ & 214008.07-231130.6 \\[7.5pt] 
42  &   &          &  21:40:33.483  0.017 & $-$23:08:31.67  0.16 & 3.45 & 
   $4.7^{+3.3}_{-2.1}$ & $2.02^{+1.07}_{-1.78}$ & $-1.60^{+1.71}_{-1.93}$ & 
   $0.7^{+0.9}_{-0.3}$ & $0.4^{+0.3}_{-0.2}$ & 214033.48-230831.6 \\[7.5pt] 
43  &   &          &  21:40:09.366  0.014 & $-$23:08:50.84  0.22 & 3.52 & 
   $13.0^{+4.7}_{-3.6}$ & $0.45^{+0.78}_{-0.76}$ & $1.31^{+1.38}_{-1.18}$ & 
   $2.1^{+2.0}_{-0.9}$ & $0.9^{+0.7}_{-0.3}$ & 214009.36-230850.8 \\[7.5pt] 
44  &   &          &  21:40:24.403  0.009 & $-$23:14:20.38  0.22 & 3.59 & 
   $11.6^{+4.5}_{-3.4}$ & $>-1.93$ & $>-1.93$ & 
   $4.8^{+2.5}_{-1.4}$ & $0.4^{+0.7}_{-0.3}$ & 214024.40-231420.3 \\[7.5pt] 
45  &   &          &  21:40:16.223  0.011 & $-$23:07:18.45  0.08 & 3.73 & 
   $87.4^{+10.4}_{-9.3}$ & $0.28^{+0.26}_{-0.26}$ & $0.44^{+0.36}_{-0.36}$ & 
   $8.8^{+1.9}_{-1.2}$ & $6.5^{+1.2}_{-0.9}$ & 214016.22-230718.4 \\[7.5pt] 
46  &   &          &  21:40:21.436  0.018 & $-$23:06:50.14  0.15 & 3.95 & 
   $43.1^{+7.6}_{-6.5}$ & $2.82^{+0.90}_{-0.78}$ & $1.92^{+0.61}_{-0.57}$ & 
   $9.3^{+3.0}_{-1.8}$ & $3.3^{+1.1}_{-0.7}$ & 214021.43-230650.1 \\[7.5pt] 
47  &   &          &  21:40:21.558  0.012 & $-$23:14:49.49  0.15 & 4.04 & 
   $16.5^{+5.1}_{-4.0}$ & $1.29^{+0.84}_{-0.78}$ & $0.86^{+0.87}_{-0.82}$ & 
   $4.0^{+2.3}_{-1.1}$ & $1.4^{+0.9}_{-0.4}$ & 214021.55-231449.4 \\[7.5pt] 
48  &   &          &  21:40:20.501  0.013 & $-$23:15:05.59  0.25 & 4.32 & 
   $14.4^{+4.9}_{-3.7}$ & $0.44^{+0.74}_{-0.72}$ & $0.22^{+0.98}_{-0.96}$ & 
   $2.2^{+1.3}_{-0.6}$ & $1.8^{+0.9}_{-0.5}$ & 214020.50-231505.5 \\[7.5pt] 
49  &   &          &  21:40:03.419  0.020 & $-$23:11:34.42  0.15 & 4.37 & 
   $28.1^{+6.4}_{-5.3}$ & $0.71^{+0.52}_{-0.51}$ & $-0.38^{+0.62}_{-0.63}$ & 
   $9.2^{+4.1}_{-1.9}$ & $6.5^{+2.4}_{-1.4}$ & 214003.41-231134.4 \\[7.5pt] 
50  &   &  Ellipt., 0.27 &  21:40:37.036  0.012 & $-$23:13:37.17  0.31 & 4.44 & 
   $13.1^{+4.7}_{-3.6}$ & $1.42^{+1.01}_{-0.91}$ & $0.59^{+0.93}_{-0.88}$ &
   $3.2^{+2.1}_{-0.9}$ & $0.9^{+0.6}_{-0.3}$ & 214037.03-231337.1 \\ 
\enddata
\tablecomments{Properties of X-ray sources in M30. Energy bands for
  counts and fluxes are in keV\@. X-ray fluxes (corrected for
  photoelectric absorption of $1.7\times10^{20}~{\rm cm}^{-2}$) are in
  units \\ of $10^{-15}~\ergsc$.  See \S\S3.2,3.3 for details.  }
\tablenotetext{a}{Bright foreground star.}
\end{deluxetable}
\clearpage
\end{landscape}

\end{document}